\PassOptionsToPackage{hyphens}{url}
\documentclass[11pt, letterpaper, logo]{template}
\usepackage{multirow}       
\usepackage{listings}       
\usepackage{placeins}       
\usepackage{subcaption}     
\usepackage{makecell}       
\usepackage{soul}           

\definecolor{CSRow}{HTML}{EEF6FB}

\usepackage[authoryear, round]{natbib}
\renewcommand{\cite}{\citep}

\graphicspath{{paper/}}

\lstnewenvironment{maccode}[2][]{%
  \lstset{
    title={#1},
    basicstyle=\footnotesize\ttfamily,
    numbers=left,
    numberstyle=\footnotesize,
    xleftmargin=2em,
    frame=single,
    showstringspaces=false,
    tabsize=2,
    breaklines=true,
    columns=fullflexible,
    keepspaces=true,
    aboveskip=0pt,
    belowskip=0pt
  }%
}{}

\title{ClawSentry: A Progressive Multi-Tier Security Monitor for Safeguarding Autonomous LLM Agents}

\paperurl={https://github.com/Elroyper/ClawSentry}

\author[]{Kai Wang, Zeming Wei, BiaoJie Zeng, Chang Jin, An Wang, Xiaokun Luan,
Zhixiao Lin, Jingjing Qu, Xia Hu, Xingcheng Xu}

\date{August 6, 2026}

\begin{document}

\begin{abstract}
As large language model (LLM) agents move from conversation to executing code, reading local files, and orchestrating external tools, a single agent hijacked by a malicious third-party skill can cause data exfiltration, privilege escalation, or cascading compromise. We argue that agentic risk is \emph{progressive}: it can enter at four loci of the agent control loop---skill admission, invocation-time intent, execution-time effect, and post-action consequence---while a denied dangerous objective can reappear across surface forms, tools, or turns; existing safeguards are typically local to one lifecycle boundary or one call. Guided by this threat model, we present \textbf{ClawSentry}, an open-source, framework-agnostic security supervision gateway for agent runtimes. Before a skill package is ever executed, \emph{First-use Skill Package Review} (FSPR) audits it under a deterministic evidence floor, escalating unresolved cases to bounded read-only agentic review (locus A). At runtime, a three-tier progressive decision engine---a deterministic L1 layer, a rule-anchored L2 semantic reviewer, and a read-only L3 evidence-seeking agent---spends contextual review only on the residual ambiguity, while a session-level anti-bypass mechanism recognizes tool-switching and rephrased retries (loci B--C); a post-action path feeds high-severity evidence non-retroactively into later review (locus D). An Agent Harness Protocol (AHP) abstraction applies one policy across Codex, Claude Code, Kimi CLI, and Gemini CLI without modifying agent internals. On SkillInject with Codex/GPT-5.4, contextual ASR falls from 39.55\% to 2.61\% while contextual TSR moves only from 83.78\% to 83.05\%. Across five Work Agents on the full SkillsSafety benchmark, ClawSentry confines ASR to 9.09--15.03\% from 33.5--49.7\% unprotected, and aggregate TSR on clean skills remains 98.7\%.
\end{abstract}

\maketitle

\section{Introduction}
\label{sec:intro}

Large Language Model (LLM)-based agents now execute code, invoke system tools, and orchestrate workflows autonomously~\cite{zheng2025deepresearcher,wu2024autogen}. Once granted local filesystem, network, and command-execution access, they pose threats far beyond those of the base LLM~\cite{lupinacci2025dark,ferrag2025prompt,deng2025ai,ma2026safety}.

These risks are \emph{progressive} and \emph{cross-stage}. Risk can enter at four loci along the generic agent control loop---skill admission, invocation-time intent, execution-time effect, and post-action consequence. Independently, an unsafe objective denied in one invocation can re-enter through rewritten arguments, a sequence of individually plausible calls, or another tool surface (\S\ref{sec:threat}). Existing defenses largely act at a \emph{single} locus or evaluate calls independently, leaving two orthogonal requirements: coverage across observable lifecycle boundaries and state across attempts~\cite{wang2025agentspec,liu2026agentdog,owasp_agentic_2026}. A selective-review gateway that satisfies both requirements, and stays agnostic to the underlying agent framework, is still largely unaddressed.

Mediating those loci requires a reviewer that can read code and judge intent---competence that is dual-use and concentrated in the frontier models making offense cheap~\cite{anthropic2026mythos,wang2026cybergym}, and that offense already fields in documented operations~\cite{anthropic2025espionage}. Such competence must enter the defense loop, yet a model granted execution authority is itself a high-value entry point. We therefore position it as an \emph{external supervisor} under three constraints---bounded exposure, no authority to act, and extensible priors---developed in \S\ref{sec:position}.

\begin{figure*}[t]
\centering
\includegraphics[width=\linewidth]{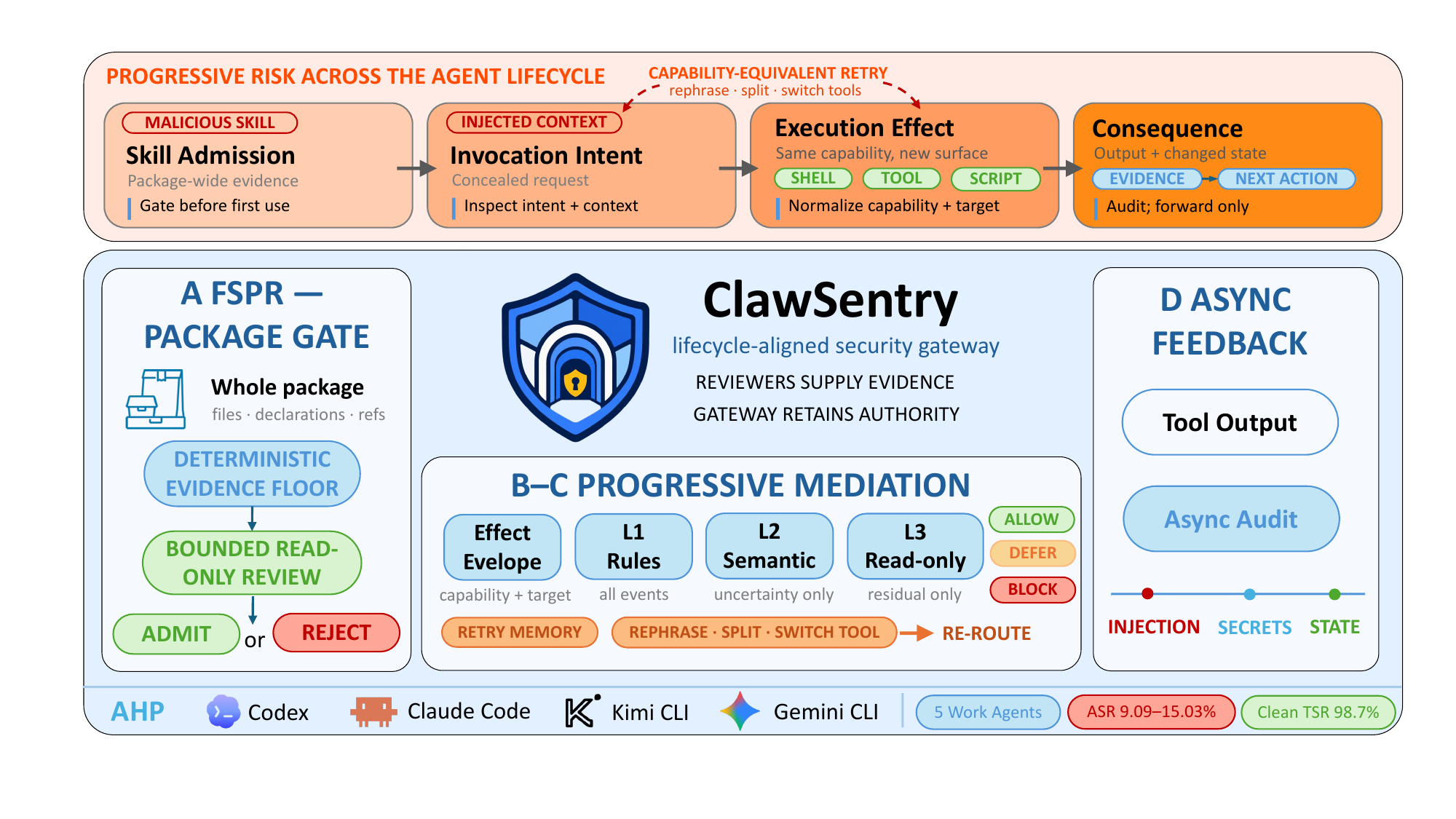}
\caption{ClawSentry aligns one gateway with four lifecycle loci. FSPR gates A;
effect-normalized L1--L3 mediation covers B--C; retry memory catches
capability-equivalent bypasses; and asynchronous D analysis informs later
actions. AHP applies one policy across four frameworks; bottom badges report
selective review and safety--utility.}
\label{fig:overview}
\end{figure*}

To address both gaps, we present \textbf{ClawSentry}, a unified security supervision gateway for AI agent runtimes built on the Agent Harness Protocol (AHP; Figure~\ref{fig:overview}). Its first line of defense is \textbf{First-use Skill Package Review} (FSPR): before a skill package is ever executed, FSPR performs a read-only, \emph{package-level} evidence audit, intercepting a malicious skill prior to use (locus A). FSPR is not a single LLM verdict: deterministic rules establish a non-negotiable evidence floor, unresolved packages receive bounded \texttt{agentic-readonly} cross-file review, and the gateway retains policy authority. Beyond pre-use review, ClawSentry routes runtime actions through a \textbf{three-tier progressive decision engine}---a deterministic \textbf{L1 Policy Engine}, a rule-anchored \textbf{L2 Semantic Analyzer}, and a read-only multi-turn \textbf{L3 Agent Analyzer} (loci B/C)---spending contextual review only on the residual ambiguity, and a \textbf{session-level anti-bypass} mechanism that records previously blocked effects and recognizes tool-switching or rephrased retries. Reviewer toolkits are read-only by construction, while versioned YAML rules and review skills let operator security priors extend their reach without widening their authority. A six-dimensional (D1--D6) risk snapshot informs tier routing, and a lightweight post-action layer asynchronously audits completed effects (locus D), letting the operator halt a session before further loss and feeding evidence into the gateway's adjudication of later actions (\S\ref{sec:threat:scope}).

During evaluation, ClawSentry integrates with Codex, Claude Code, Kimi CLI, and Gemini CLI. Across five Work Agents on the full SkillsSafety benchmark~\cite{jin2026skillsafetybench} it confines ASR to \textbf{9.09--15.03\%}, against 33.5--49.7\% raw, while preserving an aggregate \textbf{98.7\%} TSR on clean skills; on SkillInject~\cite{schmotz2026skillinject} with Codex/GPT-5.4 contextual ASR falls from 39.55\% to \textbf{2.61\%} with contextual TSR held within 0.73 points. Our principal contributions are:
\begin{enumerate}
\item A \textbf{progressive lifecycle threat model} organizing skill-execution risk into four loci plus a cross-cutting capability-equivalent bypass property, exposing the gaps a single-stage defense leaves.
\item An \textbf{open-source, framework-agnostic security supervision gateway} that decouples security policy from agent libraries via AHP, implemented across four frameworks.\footnote{Code, configuration, and review rules are available at \url{https://github.com/Elroyper/ClawSentry} under the MIT license.}
\item A \textbf{method} combining FSPR with a three-tier progressive engine and a session-level anti-bypass mechanism, positioning a capable reviewer as a read-only external supervisor.
\item An \textbf{empirical evaluation} spanning both benchmarks, five Work Agents, a clean-utility suite, a mixed package corpus, matched baselines, ablations, reviewer substitutions, and request-volume analysis.
\end{enumerate}

\section{Related Work}
\label{sec:related}

\paragraph{Autonomous LLM agents.}
LLMs have evolved into autonomous agents that plan and act via iterative reasoning frameworks~\cite{yao2023react}, accessible through open-source systems such as AutoGPT~\cite{autogpt2023} and OpenHands~\cite{wang2025openhands}, often communicating via standardized protocols like MCP~\cite{anthropic2024mcp}. Granting these agents direct access to file systems, APIs, and terminals fundamentally expands their attack surface.

\paragraph{Security risks.}
Autonomous agents are highly susceptible to \emph{indirect prompt injections}~\cite{greshake2023indirect, zhan2024injecagent}, through which malicious instructions concealed in documents or web content hijack agent execution, enabling unauthorized shell commands, privilege escalation, or data exfiltration~\cite{lupinacci2025dark}. Because these exploits are semantic rather than signature-based, traditional detection is largely ineffective~\cite{fang2024llmagents}.

\subsection{Safeguarding LLM Agents}
To mitigate the emerging threats against LLM agents, researchers have proposed various defensive layers. Early safeguards primarily focused on static input-output moderation, employing auxiliary models or rule-based toolkits to filter malicious content during human-AI conversations~\cite{inan2023llamaguard, shieldgemma2024, ghosh2024aegis, rebedea2023nemo}. For direct and indirect prompt injections, techniques such as Spotlighting~\cite{hines2024spotlighting}, heuristic self-hardening~\cite{protectai2023rebuff}, and benchmarking-driven robustness enhancements~\cite{yi2023bipia} have been developed to preserve instruction integrity.

However, as agents gain autonomous execution capabilities, prompt-level filtering proves insufficient. Recent works have introduced system-level architectures that enforce execution isolation~\cite{wu2025isolategpt}, dynamic rule-based monitoring~\cite{li2025drift}, and comprehensive security design patterns~\cite{beurerkellner2025designpatterns}. These agentic defenses often draw inspiration from traditional access control paradigms, such as Role-Based Access Control (RBAC)~\cite{sandhu1996rbac}, to reliably restrict tool privileges. Concurrently, specialized sandbox environments like ToolEmu~\cite{ruan2024toolemu} and AgentDojo~\cite{debenedetti2024agentdojo} have emerged to safely evaluate these attacks and defenses.

\paragraph{Positioning.} Prior defenses concentrate on a single locus: prompt filtering at invocation (B)~\cite{hines2024spotlighting,protectai2023rebuff}; execution isolation at effect (C)~\cite{wu2025isolategpt}; output moderation at consequence (D)~\cite{inan2023llamaguard,shieldgemma2024}. Among the representative systems summarized in Table~\ref{tab:related}, ClawSentry is the only one to combine skill admission (A) with progressive B/C runtime gating, cross-attempt migration handling, framework-agnostic integration, and end-to-end agent evaluation.

\begin{table}[htbp]
\centering
\begin{tabular}{@{}lccccc@{}}
\toprule
\textbf{System} & \textbf{A} & \textbf{B/C} & \textbf{Anti-} & \textbf{Frw.-} & \textbf{Agent} \\
 & \textbf{(skill)} & \textbf{(gate)} & \textbf{bypass} & \textbf{agn.} & \textbf{bench.} \\
\midrule
LlamaGuard~\cite{inan2023llamaguard}      & \ding{55} & \ding{55}  & \ding{55} & \checkmark & \ding{55} \\
Spotlighting~\cite{hines2024spotlighting} & \ding{55} & B only     & \ding{55} & \checkmark & \ding{55} \\
Rebuff~\cite{protectai2023rebuff}         & \ding{55} & B only     & \ding{55} & \checkmark & \ding{55} \\
IsolateGPT~\cite{wu2025isolategpt}        & \ding{55} & C only     & \ding{55} & \ding{55}  & \ding{55} \\
DRIFT~\cite{li2025drift}                  & \ding{55} & \checkmark & \ding{55} & \ding{55}  & \checkmark \\
AgentSpec~\cite{wang2025agentspec}        & \ding{55} & \checkmark & \ding{55} & \ding{55}  & \checkmark \\
\midrule
\textbf{ClawSentry}                       & \checkmark & \checkmark & \checkmark & \checkmark & \checkmark \\
\bottomrule
\end{tabular}
\caption{Coverage of representative prior agent defenses across risk loci, as reported in their respective papers. Columns: skill-admission review (A), runtime progressive gating (B/C), cross-attempt anti-bypass, framework-agnostic, and end-to-end agent benchmark. Among the systems surveyed here, ClawSentry is the only one reporting support across all axes; this is not an exhaustive comparison of the field.}
\label{tab:related}
\end{table}

 \section{Threat Model: Progressive Risks in Agent Skill Execution}
\label{sec:threat}

We organize the risks of autonomous skill execution along the \emph{generic agent control loop}, independent of any particular defense. This lens exposes the lifecycle coverage that a single-stage defense leaves open and defines the requirement evaluated end to end by ClawSentry.

\subsection{Generic Agent Skill Lifecycle}
\label{sec:threat:lifecycle}

\begin{figure}[htbp]
\centering
\includegraphics[width=\linewidth]{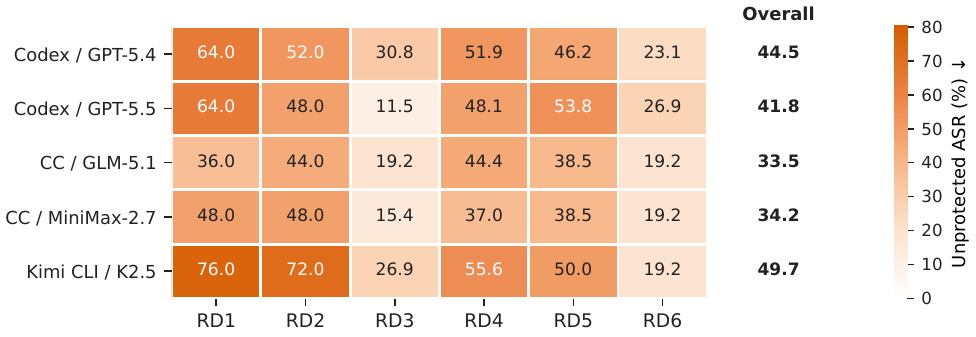}
\caption{The unprotected attack surface is structured rather than uniform.
Per-domain SkillsSafety ASR for five Work Agent configurations, as published
with the benchmark~\cite{jin2026skillsafetybench}: RD1 contextual trust and
RD2 agent permission are severe in every Work Agent family, while RD3
(execution and runtime) is markedly lower; a single aggregate rate would
average across regimes that no individual domain exhibits, which is why the
evaluation breaks results down by domain. Domain definitions are in
Appendix~\ref{app:skillsafety}.}
\label{fig:attack_surface}
\end{figure}

An autonomous agent repeatedly \emph{acquires skills}, \emph{selects a tool or command}, \emph{executes an action}, \emph{observes the output}, and \emph{updates its trajectory}~\cite{yao2023react,wang2025openhands}. Risk can enter at each boundary of the loop, and---crucially---a failure contained at one boundary can re-emerge at another as the agent iterates. We consider an attacker who either (i) publishes a malicious third-party \emph{skill package} that an agent may install and invoke (a supply-chain vector), or (ii) embeds hidden instructions in the task context the agent processes (a contextual-injection vector)~\cite{greshake2023indirect,zhan2024injecagent}. Protected assets are the host file system, credentials and secrets, network egress, and command execution. We distinguish two attack regimes used throughout the evaluation: \emph{obvious} injections, where malicious intent is explicit, and \emph{contextual} injections, where the malicious instruction is concealed within an otherwise legitimate task. This surface is measurable rather than hypothetical: Figure~\ref{fig:attack_surface} plots the per-domain attack success that the SkillsSafety benchmark reports for five unprotected Work Agent configurations.

\subsection{Four Risk Loci and One Cross-Cutting Property}
\label{sec:threat:loci}

Mapping the loop's boundaries to where risk materializes yields four risk \emph{loci} plus one property that is not tied to any single stage:

\begin{itemize}
\item \textbf{Locus A --- Skill admission / supply chain.} A third-party skill package carries malicious instructions or scripts \emph{before} it is ever invoked.
\item \textbf{Locus B --- Invocation-time intent.} At the invocation point, an action may carry direct injection, hidden intent, or obfuscated payloads~\cite{liu2023prompt}.
\item \textbf{Locus C --- Execution-time effect.} The dangerous effect of a high-risk command surfaces only when it actually runs, including across complex multi-tool context.
\item \textbf{Locus D --- Post-action consequence.} What a tool returns cannot be read off its invocation: retrieved content may carry indirect instructions that steer later steps, and the actual effect of an executed action may become visible only in its output~\cite{naik2025insecure}.
\item \textbf{Cross-cutting property --- Capability-equivalent follow-up / bypass migration.} Once an action is blocked, a determined agent can preserve its capability and target while rephrasing arguments, distributing the objective across turns, or switching tools~\cite{yan2025attack}. These transformations vary surface form and time rather than mapping one-to-one onto lifecycle loci.
\end{itemize}

\begin{figure}[htbp]
\centering
\includegraphics[width=\linewidth]{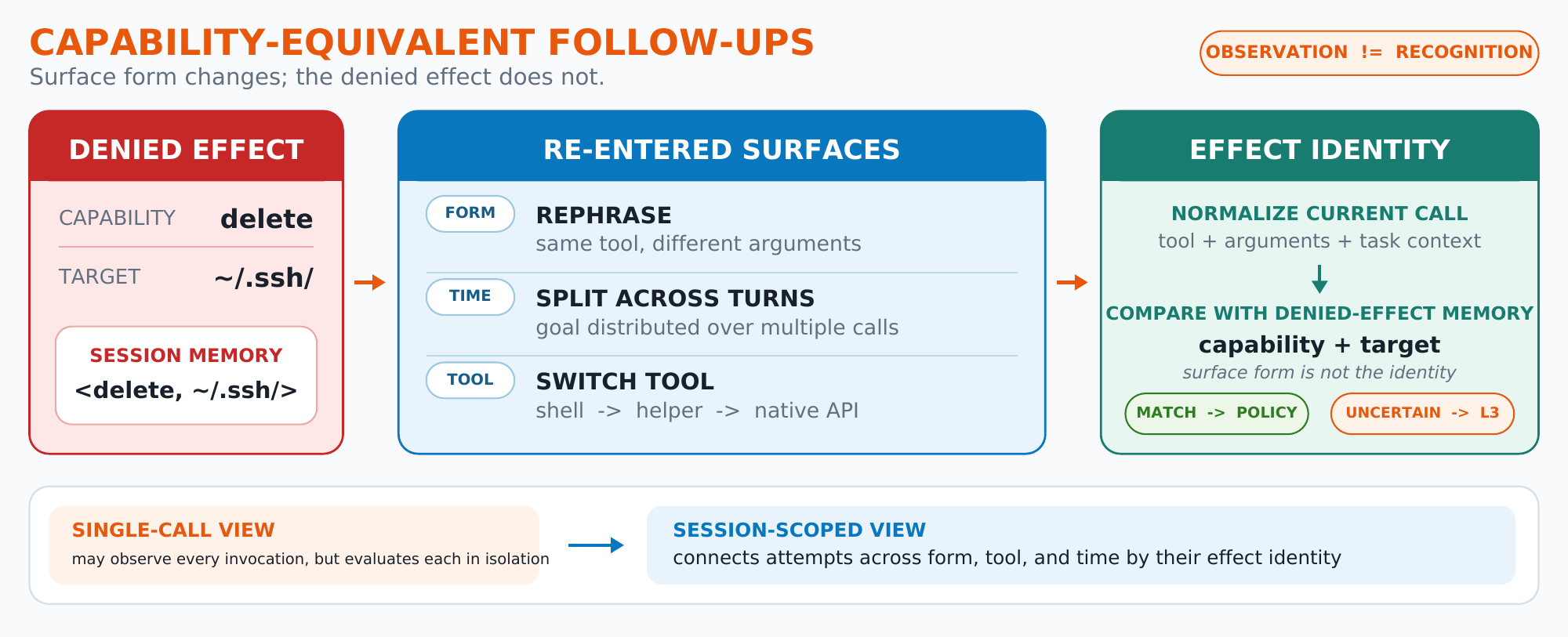}
\caption{Observation is not recognition. After a destructive effect on a
credential path is denied, the same capability and target can re-enter through
rewritten arguments to the same tool, a sequence distributed across turns, or
a different tool surface. A per-call defense may observe every resulting
invocation at its boundary while still evaluating each in isolation.
ClawSentry instead normalizes the current call by capability and target and
compares it with session-scoped denied-effect memory. High-confidence
equivalents follow the configured policy; uncertain cross-tool candidates are
routed to L3 (\S\ref{sec:method:anti_bypass}).}
\label{fig:migration}
\end{figure}

\paragraph{Why point defenses are insufficient (coverage and state requirements).} Most existing defenses act at a single locus: prompt filtering guards invocation (B), execution sandboxing guards effect (C), and output moderation guards consequence (D). Coverage of one locus is not the only limitation. A single-boundary defense may well \emph{observe} every retry form that reaches it---an invocation filter sees rephrased and tool-switched calls, and an execution sandbox constrains the same prohibited effect however the command is spelled---yet observing each attempt in isolation does not tell it that the current call repeats an objective it has already denied. Two orthogonal requirements follow: a defense must both mediate risk across the loop's observable boundaries and relate calls across surface form, tool, and time by their effect identity. \S\ref{sec:experiment} tests these requirements through end-to-end ASR, clean utility, and component evidence. Figure~\ref{fig:migration} isolates the second requirement.

\subsection{Scope and Coverage Boundaries}
\label{sec:threat:scope}

\paragraph{Scope.} We consider risks that cross an observable skill- or
runtime-mediation boundary: malicious or inconsistent packages before first
use (locus A), unsafe tool-mediated effects arising from an invocation or its
accumulated context (loci B--C), capability-equivalent follow-ups that
re-attempt a denied effect (cross-cutting), and harmful consequences that
surface only in tool output and can steer subsequent actions (locus D).
ClawSentry actively mediates loci A--C and cross-attempt migration; locus D
is observed non-retroactively. Out of scope are threats that never cross such
a boundary---direct user-to-model jailbreaks, base-model misalignment without
a tool-mediated effect, a compromised host OS, model-API manipulation, and
weight poisoning.

\section{Design Position: Frontier Security Capability as an External Supervisor}
\label{sec:position}
\label{app:discussion:asymmetry}

The competence this work asks a reviewer to supply---reading code, following
behavior across files, and judging whether an action serves the stated
task---is the same competence that finds and weaponizes vulnerabilities.  The
published technical evaluation accompanying one of the most cyber-capable
frontier models to date reports that capability as arising not from explicit
offensive training but as a downstream consequence of general improvements in
code, reasoning, and autonomy~\cite{anthropic2026mythosred}.  Offensive
and defensive security capability are therefore not two capabilities but one,
and it arrives along the general model-capability curve whether or not anyone
targets it.  Benchmarks make that curve legible: CyberGym scores agents on
reproducing 1{,}507 real vulnerabilities across 188 projects and has itself
surfaced 34 zero-days~\cite{wang2026cybergym}.

The attack side has already converted the capability into operations, as far as
public threat reporting shows.  One such report describes a single operator
using a coding agent for reconnaissance, credential harvesting, network
penetration, and ransom-demand generation against at least seventeen
organizations~\cite{anthropic2025threatintel}.  A later reported campaign drove
coding-agent instances as autonomous penetration-testing orchestrators bound to
open-source tooling through a tool protocol, with the reporting estimating
80--90\% of tactical operations executed without human action against roughly
thirty organizations~\cite{anthropic2025espionage}.  Another provider's threat
reporting describes AI as one component of a larger tool-chain spanning multiple
platforms and models rather than an isolated
capability~\cite{openai2026disrupting}.  The resulting asymmetry is
structural rather than incidental: an attacker obtains agent-level autonomy by
renting it, while defense is still largely organized around human review and
single-boundary pattern rules.

Two further observations shape what a defensive answer should look like.
First, orchestration appears to beat scale: a multi-model system coordinating
over a hundred specialized agents is reported to reach 88.4\% on CyberGym
against 83.1\% for the strongest single frontier model in the same
evaluation~\cite{microsoft2026mdash}, so the defensive
counterpart of an AI attacker is a structured system rather than one stronger
reviewer.  That result concerns vulnerability discovery; ClawSentry is its
counterpart at the runtime-mediation layer, where the object under review is an
agent's next action rather than a codebase.  Second, the capability that makes
a reviewer useful is what makes it a liability: a model competent enough to
audit is competent enough to be a high-value entry point.  The most capable
system in this class was released only as a limited preview rather than
generally~\cite{anthropic2026mythos}, which is consistent with treating such
capability as inherently dual-use.  Frontier security capability must enter the
defense loop, but not as the entity that acts.
Table~\ref{tab:supervisor_constraints} states the three constraints this
position imposes together with the mechanism each produces and where the paper
establishes it; its final row records the corollary of the third, that priors
extensible by the operator also imply supervision that is not bound to one model
family.

\begin{table}[htbp]
\centering
\footnotesize
\setlength{\tabcolsep}{3pt}
\begin{tabular}{@{}>{\raggedright\arraybackslash}p{0.30\columnwidth}>{\raggedright\arraybackslash}p{0.34\columnwidth}>{\raggedright\arraybackslash}p{0.28\columnwidth}@{}}
\toprule
\textbf{Constraint} & \textbf{Mechanism} & \textbf{Evidence} \\
\midrule
Bounded exposure: cost and latency forbid reviewing every event &
L1$\rightarrow$L2$\rightarrow$L3 selective-escalation funnel &
13.8\% of L1 volume; L3 alone 1.37\% (\S\ref{sec:exp:cost}) \\
\addlinespace
No authority to act: the reviewer is itself an attack surface and must never become an actuator &
Read-only toolkits for FSPR and L3; analyzers supply evidence, the gateway
holds policy authority &
\S\ref{sec:method:fspr}, \S\ref{sec:method:progressive} \\
\addlinespace
Extensible priors: reviewer competence must not be capped by one model's priors &
Versioned YAML attack-pattern rules; L3 review skills as YAML manifests
validated against the global read-only whitelist &
\S\ref{sec:method:progressive}, Appendix~\ref{app:configs} \\
\addlinespace
Supervision must not be bound to one model family &
Reviewer substitution across providers &
ASR 11.11--17.24\% over five providers (\S\ref{sec:exp:ablation}) \\
\bottomrule
\end{tabular}
\caption{Constraints on using frontier security capability defensively, the
ClawSentry mechanism each produces, and where the paper establishes it.}
\label{tab:supervisor_constraints}
\end{table}

The configuration evaluated in \S\ref{sec:experiment} uses a
cost-efficient reviewer (\texttt{gemini-3.5-flash}) rather than a frontier
security model, so the profile this paper reports does not depend on
frontier-tier review.  The funnel is what makes stronger review affordable:
\S\ref{sec:exp:cost} measures how small a fraction of L1 traffic ever reaches
model-backed review, so substituting a reviewer an order of magnitude more
expensive scales the review bill by that fraction rather than by full
tool-call traffic.  \S\ref{sec:exp:ablation} then locates where the protection
resides, by substituting reviewers across providers inside the same
gateway-owned enforcement chain.

 \section{Agent Harness Protocol (AHP)}
\label{sec:ahp}

A security gateway that must be re-implemented for every agent framework is not a gateway but a fork. The central enabling contribution of ClawSentry is therefore not any single detector but the \emph{Agent Harness Protocol} (AHP): a framework-agnostic abstraction that decouples agent-specific event streams from the safety logic, so that one policy engine supervises heterogeneous agents unchanged. Every mechanism in \S\ref{sec:methodology}---FSPR, the three-tier engine, anti-bypass, and post-action observation---consumes AHP events rather than framework-native ones, which is what lets ClawSentry attach to four mainstream frameworks (Codex, Claude Code, Kimi CLI, Gemini CLI) without modifying agent internals.

\paragraph{Canonical event schema.} Prior surveys document heterogeneous communication mechanisms and fragmented protocols across LLM-based agent systems~\cite{kong2025survey,yan2025beyond}. To give the security gateway a stable boundary across those interfaces, we define AHP around a unified \texttt{CanonicalEvent} schema and a standard adapter interface. Each framework-specific adapter normalizes heterogeneous native events---pre-action hooks, tool executions, and post-action results---into AHP events with strict fields (a SHA-256 \texttt{event\_id}, \texttt{session\_id}, \texttt{tool\_name}, and a generic \texttt{payload}), preserving framework-specific state in an isolated \texttt{framework\_meta} structure. Because the safety core sees only canonical events, adding a new framework requires only a lightweight adapter and inherits the full safety evaluation without touching the gateway.

\begin{figure}[htbp]
\centering
\includegraphics[width=\linewidth]{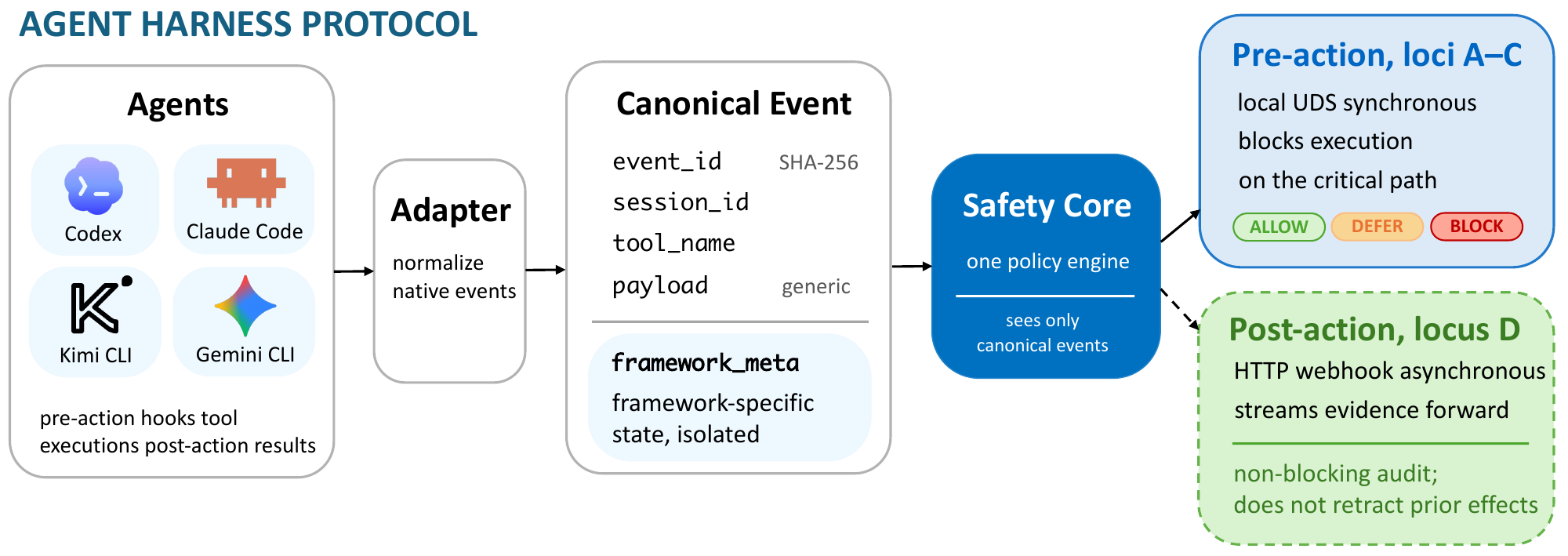}
\caption{AHP places one policy engine behind four heterogeneous event streams.
Each adapter normalizes its framework's pre-action hooks, tool executions, and
post-action results into a \texttt{CanonicalEvent}, keeping what does not
generalize in an isolated \texttt{framework\_meta} field, so the safety core
sees only canonical events. Pre-action events (loci A--C) travel over a
synchronous local UDS that blocks execution until the gateway returns Allow,
Defer, or Block; post-action events (locus D) stream asynchronously over HTTP
webhooks
for non-blocking audit. Adding a framework costs an adapter and nothing in the
gateway.}
\label{fig:ahp}
\end{figure}

\paragraph{Dual-channel mediation.} This abstraction establishes a dual-channel model matched to the risk loci of \S\ref{sec:threat}. For \emph{pre-action} events (loci A--C), the adapter synchronously blocks execution over a local Unix Domain Socket (UDS) until the gateway returns a decision (Allow, Block, or Defer), keeping enforcement on the critical path. For \emph{post-action} events (locus D), it streams results asynchronously via HTTP Webhooks for non-blocking auditing. The two channels together give ClawSentry a single, framework-independent vantage point over the entire agent control loop. Figure~\ref{fig:ahp} shows the schema and both channels together.

\section{Methodology}
\label{sec:methodology}

ClawSentry organizes supervision around the agent lifecycle rather than as a
uniform per-call filter.  Every mechanism described below consumes the
canonical AHP events of \S\ref{sec:ahp}, each paired with gateway-owned
package, workspace, and trust context, so adding a framework changes the
adapter and not the admission or runtime policy.  Throughout this path,
deterministic hard evidence is monotone, analyzers supply evidence rather than
policy actions, and the gateway retains final authority.

\subsection{Hybrid First-Use Skill Package Review}
\label{sec:method:fspr}

A runtime guard sees an action only after the skill that shaped it has
already influenced the agent, and behavior distributed across package files
cannot be recovered from a single call.  FSPR therefore reviews a package
\emph{before its first capability-bearing action}, in three stages.

\smallskip\noindent\textbf{Deterministic floor.}  A scanner builds a bounded
inventory of declarations, files, entry points, references, and observed
capabilities, and rules convert concrete inconsistencies and suspicious
structures into an \emph{evidence floor}: high-severity findings cannot be
erased by a later model response and short-circuit review when they already
establish the admission risk.  Evidence is organized along package-specific
axes---identity integrity, capability--manifest alignment, data-boundary and
execution-surface control, instruction-channel integrity, state-mutation
scope, re-entry and persistent activation, and review-evidence quality.

\smallskip\noindent\textbf{Agentic review.}  Unsettled cases enter an
\texttt{agentic-readonly} reviewer that can list, read, and search within the
skill root to connect cross-file behavior, while its toolkit exposes no
write, execution, installation, or network action.  The reviewer returns a
validated evidence record, which the gateway merges with the deterministic
floor before deciding admission: rules hold authority over known invariants,
and agentic inspection covers relationships that single-file patterns cannot
express.

\smallskip\noindent\textbf{Decision and caching.}  Reviews are cached under
an identity derived from package content and review-policy versions; a
changed package or policy requires a fresh decision.  In the evaluated
profile, policy-eligible \emph{suspicious} and \emph{inconsistent} verdicts
block admission.  FSPR is distinct from runtime L3
(\S\ref{sec:method:progressive}): it judges package-wide evidence before use,
whereas L3 investigates one ambiguous action in its session context.

\subsection{Effect-Normalized Progressive Adjudication}
\label{sec:method:scoring}
\label{sec:method:progressive}

Once a package is admitted, the same effect may appear as a shell command, a
helper script, or a framework-native tool call.  ClawSentry therefore maps
each event to an effect envelope containing the requested capability, target,
scope, artifact role, and evidence provenance.  The resulting snapshot uses
six dimensions: tool danger \(D_1\), target sensitivity \(D_2\), command
danger \(D_3\), accumulated session risk \(D_4\), agent trust risk \(D_5\),
and injection evidence \(D_6\).  Its composite score is
\begin{equation}
  \begin{aligned}
  B_t &= w_{123}\max(D_1,D_2,D_3)+w_4D_4+w_5D_5,\\
  S_t &= B_t\left(1+\lambda D_6/3\right).
  \end{aligned}
  \label{eq:risk-score}
\end{equation}

\begin{table}[htbp]
\centering
\begin{tabularx}{\linewidth}{@{}>{\raggedright\arraybackslash}p{3.6cm}>{\raggedright\arraybackslash}X>{\raggedright\arraybackslash}X@{}}
\toprule
\textbf{Dimension} & \textbf{Evaluation Focus} & \textbf{High-Risk Example} \\ \midrule
\textbf{D1: Tool-Type Danger} & Intrinsic capability of the invoked tool. & A shell or execution tool operating on a system resource. \\ \midrule
\textbf{D2: Target Sensitivity} & Security role of the normalized target path or resource. & Accessing \texttt{/etc/shadow}, \texttt{.ssh/}, or \texttt{.env}. \\ \midrule
\textbf{D3: Command-Pattern Danger} & Destructive or high-impact structure within a shell/exec command. & Recursive deletion, privilege modification, or an encoded \texttt{eval}. \\ \midrule
\textbf{D4: Session Accumulation} & Recent high-risk decisions in the current session. & Several high-risk actions preceding a staged archive-and-egress attempt. \\ \midrule
\textbf{D5: Agent Trust Risk} & Trust level supplied by gateway-owned agent/workspace context. & An untrusted or missing-trust execution context. \\ \midrule
\textbf{D6: Injection Signal} & Heuristic instruction-manipulation evidence, including content origin. & An external document containing hidden override instructions or zero-width controls. \\ \bottomrule
\end{tabularx}
\caption{The six detector dimensions used by the runtime risk snapshot. These
dimensions are distinct from the benchmark's RD1--RD6 attack domains.}
\label{tab:risk_scoring}
\end{table}

The maximum avoids counting several surface descriptions of one effect as
independent evidence; the injection multiplier makes steering more
consequential when the underlying action is already capable.  The score is a
shared routing and audit substrate, not the verdict: hard rules and
evidence-generated routing intents can still decide or escalate review
(dimension definitions in Table~\ref{tab:risk_scoring}; rules, weights, and
thresholds in Appendix~\ref{app:configs}).

\begin{figure}[htbp]
\centering
\includegraphics[width=0.78\linewidth,trim=35bp 0 180bp 0,clip]{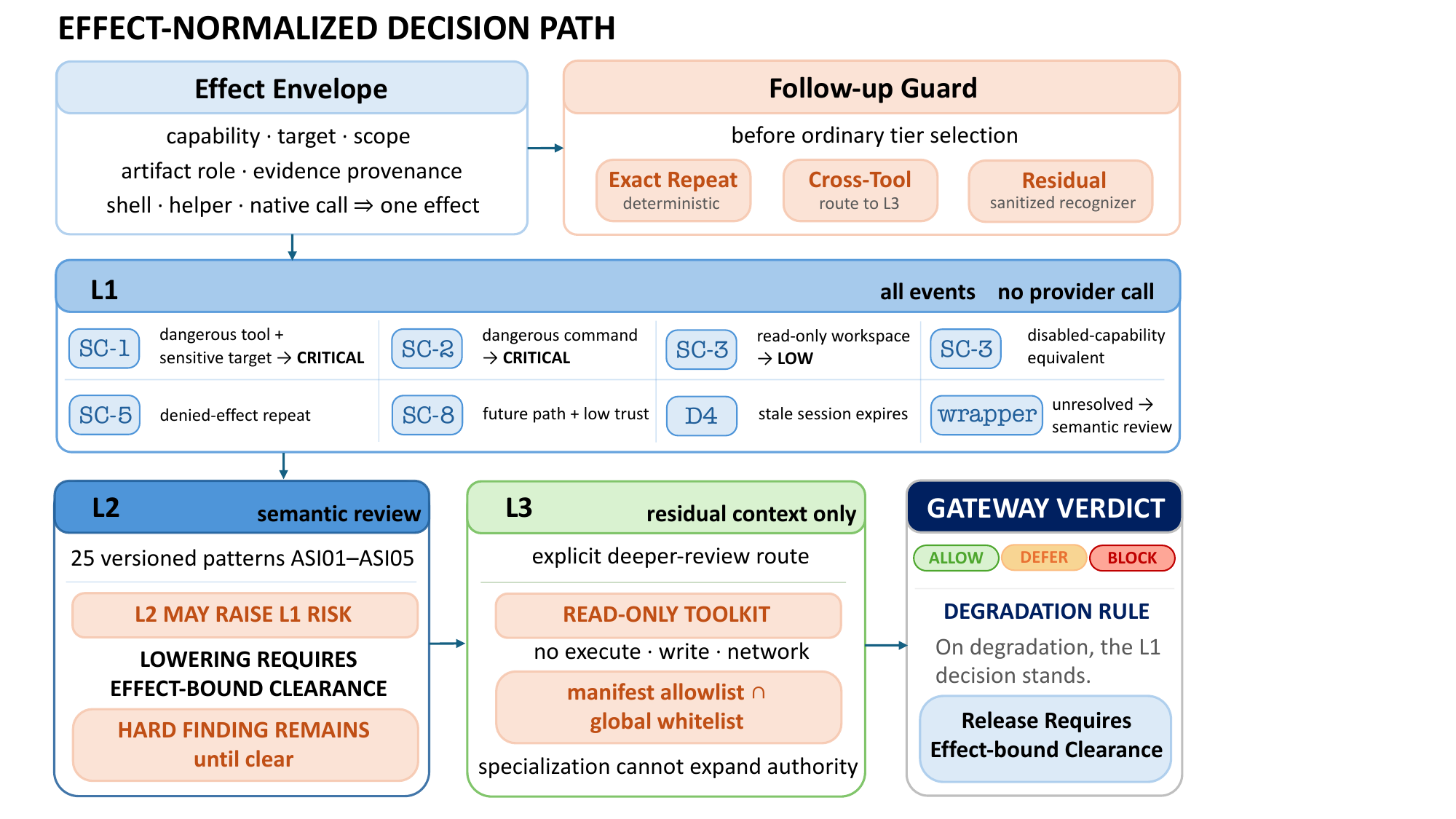}
\caption{The decision path a canonical event traverses; most events close at
L1 and only the residual escalates. Normalization to an effect envelope
precedes every other step, so one capability presented through different
surfaces is adjudicated as one effect. The follow-up guard is
consulted before ordinary tier selection, and its recognizer activates a
gateway-configured review route rather than choosing an outcome. L1 closes
clear cases through deterministic short circuits without a provider call; L2
may raise L1 risk freely but lowering it requires separately validated
effect-bound clearance, and a hard finding remains in force regardless; L3
inspects bounded evidence through a toolkit with no execute, write, or network
authority, its per-skill allowlist intersected with the gateway's global
read-only whitelist. Throughout, analyzers supply evidence and the gateway
holds policy authority.}
\label{fig:decision_flow}
\end{figure}

Review follows a selective-escalation funnel: hard evidence closes at L1,
semantic uncertainty selects L2, and only the residual context-dependent set
reaches L3.  Escalation is selective for two reasons that point the same way:
contextual review is the costly tier, and the reviewer performing it is itself
an asset whose exposure should be minimized---so an event closed at L1 is an
event no semantic reviewer ever sees, serving budget and containment at once. Figure~\ref{fig:decision_flow} traces the whole path, including the short circuits and the clearance asymmetry described below.

\textbf{L1} evaluates the normalized
snapshot and deterministic short circuits without a provider call: sliding
windows expire stale session risk, effect-aware short circuits close clear
cases, and disabled-capability equivalents, unresolved wrappers, and generated
future-execution paths are routed conservatively
(Appendix~\ref{app:configs}).

\textbf{L2} combines versioned attack-pattern rules with a single-turn LLM
for semantic intent.  The built-in library contains 25 patterns derived from
the first five categories (ASI01--ASI05) of OWASP's Top 10 for Agentic
Applications~\cite{owasp_agentic_2026},
and operators may replace the versioned YAML rule file without changing L2's
decision contract (Appendix~\ref{app:configs}).  L2 may raise L1 risk; to
lower it, L2 needs a separately validated clearance bound to the current event
and effect, and a hard finding remains in force regardless.

\textbf{L3} is invoked by an explicit deeper-review route, or when L2 is
non-decisive and the L3 trigger policy matches.  Its multi-turn analyzer may
inspect bounded workspace, trajectory, and session-risk evidence through
read-only tools whose toolkit exposes no command execution, write, or network
action; domain-specific review skills can be supplied as YAML manifests whose
tool allowlists are validated against the gateway's global read-only
whitelist, so specialization cannot expand L3's authority
(Appendix~\ref{app:configs}).  The constraint is containment as well as
policy: a reviewer that must read attacker-influenced material cannot be
turned into an actuator, and a subverted review turn costs evidence quality
rather than host state.  A decisive L2 result skips L3 unless an
explicit route requires it.  If optional semantic analysis degrades, the
established L1 decision remains in force, and an action requiring contextual
review is not released without valid effect-bound clearance.

\subsection{Operator-Granted Session Trust}
\label{sec:method:scope}

Whether an action is \emph{warranted} depends on what the session is for:
generic rules rightly treat a delivery or deployment path as dangerous, yet
some tasks legitimately need exactly those targets.  ClawSentry therefore
applies least privilege~\cite{saltzer1975protection} at session level: an optional
\emph{session scope profile} records the authority a task actually requires,
instead of letting it be inherited by default.  Grants take effect only
through operator-class channels, so text an attacker can influence cannot
widen its own authority.  Scope evaluation is monotone: it
can tighten a decision to \textsc{Defer} or \textsc{Block}, and leaves
verdicts from FSPR, the follow-up guard, and the tiers as they stand.

\subsection{Capability-Equivalent Follow-up Guard}
\label{sec:method:anti_bypass}

Independent per-call checks forget that an effect was already denied: an
agent can rename a command, rephrase it later, or switch tools while
preserving the same capability and target.  ClawSentry instead stores
compact, session-scoped records of denied effects---richer fingerprints for
finalized high-risk decisions, holds for unresolved \textsc{Defer}---and
queries them before ordinary tier selection.  Exact repeats and normalized
destructive effects are handled deterministically, while uncertain cross-tool
equivalence promotes the new event to L3; for residual candidates, a
sanitized LLM recognizer judges only whether the current action is a
follow-up to a prior high-risk blocked or deferred action.  The recognizer
cannot choose the enforcement outcome: a positive match activates the
gateway-configured L3 review route, which retains final authority.  This
separates \emph{semantic recognition} from \emph{semantic authority}.

\subsection{Post-Action Consequence Feedback}
\label{sec:method:post_action}

Some evidence appears only in tool output, after the action that produced it
has completed.  At locus D the effect has already taken place, and mainstream
harnesses expose no hook that retracts or rewrites a returned tool result, so
this stage is where ClawSentry reports rather than enforces.  It
asynchronously analyzes \textsc{post-action} output for indirect
instructions, secret exposure, exfiltration, and obfuscation; the completed
event itself stays \textsc{Allow}.  Eligible findings are broadcast as
evidence, so the operator learns what a tool actually returned---or what an
executed action actually did---and can stop the task before the agent builds
on it.  A high-severity result may also raise scrutiny on a subsequent action
if analysis completes first, adding bounded forward feedback.

\section{Experiments}
\label{sec:experiment}

The evaluation tests ClawSentry's central deployment claim: one
framework-agnostic gateway can create a strong asymmetry between toxic and
clean skill use. Across five Work Agents, ClawSentry contracts SkillsSafety ASR
from 33.5--49.7\% raw to 9.09--15.03\%, while preserving 98.7\% aggregate TSR
on clean skills. We establish this security--utility profile, then account for
the task-success cost it carries on poisoned packages---where that cost falls
and what it buys---before isolating the mechanisms that produce protection and
measuring reviewer portability and selective-review volume.

\subsection{Experimental Setup}
\label{sec:exp:setup}

\paragraph{Frameworks and Work Agents.}
ClawSentry attaches to Codex, Claude Code, Kimi CLI, and Gemini CLI through
AHP. The end-to-end evaluation uses five Work
Agent configurations---Codex with GPT-5.4~\cite{openai2026gpt54} and
GPT-5.5~\cite{openai2026gpt55}, Claude Code with
GLM-5.1~\cite{glm5team2026glm5vibecodingagentic} and MiniMax-2.7
(\texttt{minimax-m2.7})~\cite{chen2026minimaxm2seriesminiactivations}, and
Kimi CLI with Kimi-K2.5~\cite{kimiteam2026kimik25visualagentic}---where
``Work Agent'' always denotes the task-executing model, distinct from the
ClawSentry safety reviewer. The controlled studies of
\S\ref{sec:exp:ablation} instead run Claude Code with MiniMax-M3
(\texttt{minimax-m3})~\cite{lai2026minimaxsparseattention}, a separate MiniMax
release from the MiniMax-2.7 of the main experiments; each of those studies
varies a single factor inside its own fixture, so its values compare within a
block and are not pooled with the main tables.

\paragraph{Reviewer configuration.}
Unless a reviewer-substitution experiment states otherwise, every protected
main-experiment condition uses
\texttt{gemini-3.5-flash}~\cite{google2026gemini35flash} as the safety
reviewer, serving two independent paths: Gateway L2/L3 reviews runtime actions,
while FSPR reviews a skill package at first use.

\paragraph{Benchmarks.}
We evaluate on two public skill-security benchmarks and two auxiliary suites
constructed from their material.
\textbf{SkillInject}~\cite{schmotz2026skillinject} contains 180 obvious
injections with explicit malicious intent and 139 contextual injections
embedded in legitimate tasks.
\textbf{SkillsSafety}~\cite{jin2026skillsafetybench} contains 155 cases across
six risk domains and evaluates both unsafe behavior and completion of the
benign user task; its benign task carriers are drawn from
\textbf{SkillsBench}~\cite{li2026skillsbench}. The
\textbf{clean-utility} suite contains 58 clean tasks with clean skills drawn
from SkillInject; the \textbf{101-package mixed corpus} isolates FSPR
discrimination using 48 toxic SkillsSafety packages, 48 clean SkillsBench
packages, and five clean SkillInject regression packages. Each controlled
study uses a 60-case SkillsSafety subset drawn at random under a fixed seed and
held constant across that study's conditions; the utility analysis of
\S\ref{sec:exp:tsr_analysis} uses a separate such sample.

\paragraph{Task-artifact declarations.}
Protected SkillsSafety sessions use the task-runner channel of
\S\ref{sec:method:scope}: the harness issues per-case operator-granted
artifact declarations, so verifier-mandated locations are adjudicated as part
of the task contract (Appendix~\ref{app:scope_trust}).

\paragraph{Metrics and evaluation convention.}
Attack success rate (ASR, lower-is-better) and task success rate (TSR, the
completion of the benign user task, higher-is-better) are the metrics defined
by the two benchmarks, and we adopt each benchmark's own criteria and
evaluation harness unmodified rather than introducing a scoring rule of our
own: SkillsSafety judges unsafe behavior with a deterministic verifier and
scores benign-task completion from the case's verifier
tests~\cite{jin2026skillsafetybench}, whereas SkillInject scores both attack
success and task completion with its bundled LLM
judge~\cite{schmotz2026skillinject}. Within every raw/protected pair the two
conditions are scored by the same harness, so what a protected run is credited
or penalized for is fixed by the benchmark rather than by us.
TSR uses one strict binary criterion throughout: only a full
verifier pass counts, partial credit never does. Runs interrupted by
infrastructure faults (\emph{technical} outcomes) are rerun where possible
and otherwise excluded from every denominator. \S\ref{sec:exp:tsr_analysis}
keeps the same criterion but fixes the denominator at its 60 sampled cases;
this fixed-denominator view is not pooled with the benchmark-specific TSR
elsewhere. ClawSentry natively returns \textsc{Allow}, \textsc{Defer}, or
\textsc{Block}; the unattended benchmark deterministically enforces
\textsc{Defer} as \textsc{Block}, a safety-first operating point from which
interactive review can recover approved boundary actions
(Appendix~\ref{app:defer}).

\subsection{Security Effectiveness on Toxic Skills}
\label{sec:exp:security}

\begin{figure}[htbp]
\centering
\includegraphics[width=\linewidth]{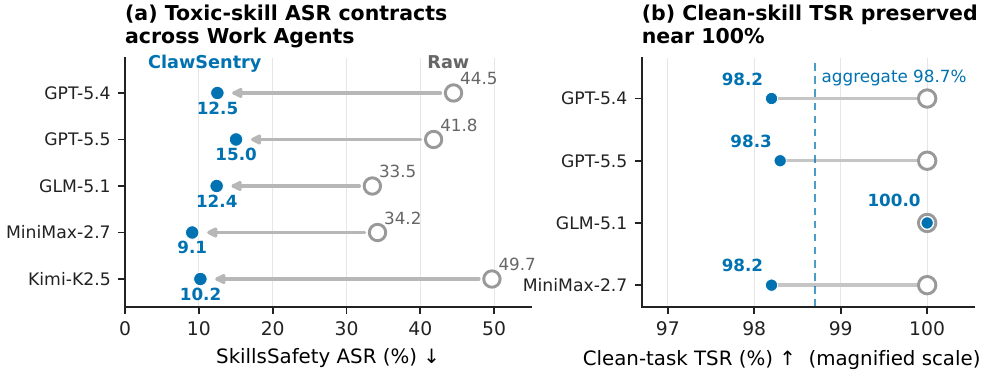}
\caption{The security--utility asymmetry the gateway is built to produce.
(a) SkillsSafety ASR for five Work Agents, unprotected (open) versus protected
(filled), with exact values at both endpoints; full results are in
Table~\ref{tab:skillsafety_main}, denominators in
Appendix~\ref{app:skillsafety}. (b) Clean-skill task
success on 58 clean tasks per Work Agent: raw TSR is 100\% for all four
configurations and protected TSR remains at 98.2--100\%, an aggregate of
98.7\%. The horizontal scale in (b) is magnified.}
\label{fig:asymmetry}
\end{figure}

\paragraph{SkillInject.}
Table~\ref{tab:skillinject} pairs each Work Agent with its protected
counterpart. The raw rows establish an attack surface spanning Work Agent
families---obvious ASR 18.75--35.26\% and contextual ASR 27.74--53.91\%---and
the protected rows suppress both direct and concealed attacks in every
configuration: obvious ASR contracts to \textbf{4.09--6.02\%}
(4.2--7.2$\times$), contextual ASR to \textbf{2.61--14.29\%}
(3.5--15.2$\times$). The legitimate contextual task survives on Codex, whose
protected TSR stays within 0.73 and 2.26 points of raw.

\begin{table}[htbp]
\centering
\small
\setlength{\tabcolsep}{3.5pt}
\begin{tabular}{@{}llccc@{}}
\toprule
\textbf{Work Agent} & \textbf{Cond.} &
\textbf{Obv.\ ASR}$\downarrow$ &
\textbf{Ctx.\ ASR}$\downarrow$ &
\textbf{Ctx.\ TSR}$\uparrow$ \\
\midrule
& Raw & 18.75 & 39.55 & 83.78 \\
\rowcolor{CSRow}\multirow{-2}{*}{Codex/GPT-5.4} &
\textbf{CS} & 4.47 & \textbf{2.61} & 83.05 \\
& Raw & 29.38 & 39.86 & 87.01 \\
\rowcolor{CSRow}\multirow{-2}{*}{Codex/GPT-5.5} &
\textbf{CS} & \textbf{4.09} & 8.40 & \textbf{84.75} \\
\midrule
& Raw & 31.10 & 53.91 & 81.94 \\
\rowcolor{CSRow}\multirow{-2}{*}{CC/GLM-5.1} &
\textbf{CS} & 4.49 & 14.29 & 76.27 \\
& Raw & 35.26 & 37.60 & 83.33 \\
\rowcolor{CSRow}\multirow{-2}{*}{CC/MiniMax-2.7} &
\textbf{CS} & 6.02 & 10.85 & 69.49 \\
\midrule
& Raw & 34.66 & 27.74 & 79.22 \\
\rowcolor{CSRow}\multirow{-2}{*}{Kimi CLI/K2.5} &
\textbf{CS} & 5.63 & 6.82 & 64.41 \\
\bottomrule
\end{tabular}
\caption{SkillInject results (\%); CS denotes ClawSentry, CC Claude Code, and
light-blue rows mark protected conditions. Bold marks the best protected value
per column. The 180-case obvious subset has
no legitimate-task objective; the 139-case contextual subset reports both ASR
and TSR.}
\label{tab:skillinject}
\end{table}

\paragraph{SkillsSafety.}
Table~\ref{tab:skillsafety_main} shows an ASR reduction for every Work
Agent: the raw 33.5--49.7\% range contracts to \textbf{9.09--15.03\%}
(Figure~\ref{fig:asymmetry}(a)), a 2.7--4.9$\times$ reduction and an absolute
improvement of 21.1--39.5 points.
Raw values are the benchmark's published baselines for these same five
configurations~\cite{jin2026skillsafetybench}; the protected values are ours,
all under one reviewer configuration, so the breadth reflects transfer across
Work Agents rather than reviewer changes. Auditable
counts appear in Appendix~\ref{app:skillsafety}.

Task success moves in the opposite direction: protected TSR falls by 5.0--14.7
points. This raises the question of whether the gateway obtains safety at the
expense of completing legitimate work, which the next two subsections address
from the two sides that matter in deployment: \S\ref{sec:exp:utility} evaluates
clean tasks with clean skills, the regime in which an agent operates most of the
time, and \S\ref{sec:exp:tsr_analysis} decomposes the poisoned sessions measured
here, for which both TSR columns contain the poisoned package.

\begin{table}[htbp]
\centering
\footnotesize
\setlength{\tabcolsep}{4.2pt}
\begin{tabular}{@{}lcccc@{}}
\toprule
\textbf{Work Agent} &
\multicolumn{2}{c}{\textbf{ASR}$\downarrow$} &
\multicolumn{2}{c}{\textbf{TSR}$\uparrow$} \\
\cmidrule(lr){2-3}\cmidrule(lr){4-5}
& Raw & +CS & Raw & +CS \\
\midrule
Codex/GPT-5.4 & 44.5 & 12.50 & 38.7 & 25.66 \\
Codex/GPT-5.5 & 41.8 & 15.03 & 42.6 & 30.07 \\
\midrule
CC/GLM-5.1 & 33.5 & 12.40 & 37.4 & 23.14 \\
CC/MiniMax-2.7 & 34.2 & \textbf{9.09} & 20.6 & 15.58 \\
\midrule
Kimi CLI/K2.5 & 49.7 & 10.20 & 31.0 & 16.33 \\
\bottomrule
\end{tabular}
\caption{SkillsSafety aggregate results (\%); CC denotes Claude Code. Raw
columns are the published benchmark
baselines~\cite{jin2026skillsafetybench}. Both TSR columns are measured with the
poisoned package present; \S\ref{sec:exp:tsr_analysis} decomposes that operating
point.}
\label{tab:skillsafety_main}
\end{table}

\subsection{Impact on Clean Work}
\label{sec:exp:utility}

We first quantify what the policy costs when nothing is wrong. The clean-utility
suite holds task and skill both clean, so any TSR it gives up
is pure policy overhead. Raw TSR is 100\% for all four Codex/Claude Code
configurations; with ClawSentry, each pair remains at 98.2--100\%, and the
aggregate is \textbf{98.7\%} (Figure~\ref{fig:asymmetry}(b)). The result is
unambiguous on this side: protection is not obtained by refusing ordinary work,
and taken together with the toxic-skill ASR reductions above it establishes the
intended asymmetry---the gateway preserves ordinary skill-assisted work while
imposing a strong boundary on compromised packages. It also localizes the
remaining question, since a 1.3-point overhead cannot account for a 5.0--14.7
point margin: that margin arises entirely in sessions where a poisoned package is
present, which \S\ref{sec:exp:tsr_analysis} decomposes.

\subsection{Accounting for the Poisoned-Case Margin}
\label{sec:exp:tsr_analysis}

The second side requires decomposing the margin itself. We use a four-condition
comparison on a 60-case
Codex/GPT-5.4 sample spanning RD1--RD6, each poisoned case paired with a
same-class clean SkillsBench skill: (A)~\emph{no\_skill}, the agent working
alone; (B)~\emph{clean\_skill}; (C)~\emph{poisoned\_skill}, the original
compromised package with ClawSentry disabled; and (D)~the same package under the
hard-block policy. A--C carry no ClawSentry component, which is what makes the
comparison diagnostic: poisoning is measured before any defense exists.

\begin{figure}[htbp]
\centering
\includegraphics[width=\linewidth]{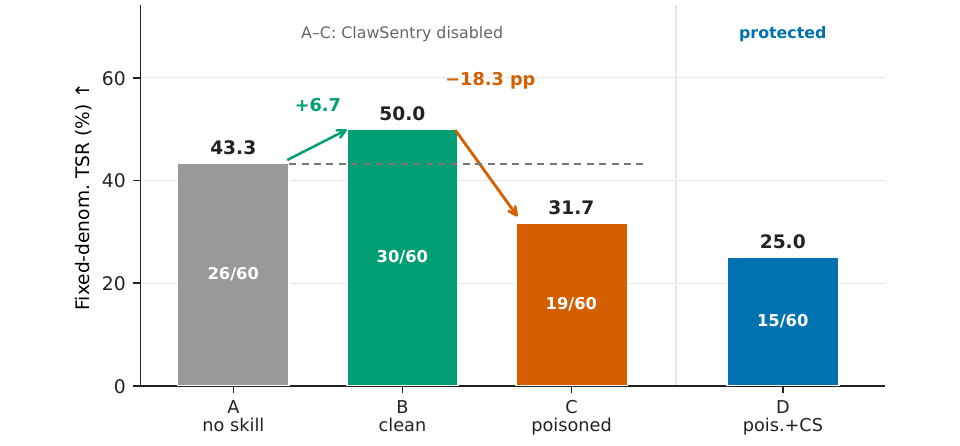}
\caption{Fixed-denominator task-success diagnostic on the 60 sampled
Codex/GPT-5.4 SkillsSafety cases; the dashed line is the no-skill baseline.
Conditions A--C contain no ClawSentry component and isolate the attack-side
effect: a clean skill adds 6.7 points over that baseline, whereas poisoning
removes 18.3 points before enforcement is introduced. D is the protected
hard-block operating point, a further 6.7 points below C. Bar interiors give
full-task-success counts.}
\label{fig:tsr_poisoning}
\end{figure}

Figure~\ref{fig:tsr_poisoning} gives the decomposition on these 60 cases, and
Figure~\ref{fig:tsr_per_domain} breaks it down by risk domain (counts in
Appendix~\ref{app:tsr_analysis}). The skill under test is
genuinely useful: a clean same-class package raises TSR from 43.3\% (A) to
50.0\% (B). Measured against that reference, and contrary to the common
assumption that a stealthy attack preserves the usefulness of its host skill,
\emph{poisoning is itself destructive}: with ClawSentry disabled the poisoned
package reaches only 31.7\% (C), 18.3 points below the clean skill and 11.7
points below the agent's own no-skill baseline. Enforcement accounts for the
remaining 6.7 points, from C to 25.0\% (D). Of the 25.0-point distance between
the clean-skill reference and the protected operating point, therefore,
approximately three quarters is incurred before the gateway makes any decision,
and that portion is already reflected in the raw column of
Table~\ref{tab:skillsafety_main}.

We attribute the remaining 6.7 points from C to D to overblocking, and report it
as a direct cost of enforcement rather than a measurement artifact. It is the
cost any enforcement boundary incurs when decisions are made without human
review, and in this setting its scale is measurable: FSPR's clean false-block
rate is 5.7--7.5\% across two independent review backends
(\S\ref{sec:exp:fspr}). The unattended harness amplifies this
cost by construction, since it resolves every \textsc{Defer} as \textsc{Block};
D is therefore a lower bound on the attended operating point
(Appendix~\ref{app:defer}). Two structural properties of the setting bound how
small the remainder can be. First, the benign task and the attack path arrive in
the same artifact, so a decision about one is necessarily a decision about both.
Second, that decision is made inside a session the package has already steered,
so refusing an action truncates the current trajectory rather than restoring the
trajectory the agent would have followed had the package never been offered.

Whether this remainder can be closed is a matter of design choice rather than a
gap in measurement. Once a poisoned package has been admitted, the attainable
reference is C=31.7\% rather than the clean-skill 50.0\%, and the remaining
points up to it are recoverable only by granting the package the latitude that
the attack is designed to exploit. A guardrail must therefore choose between
pursuing a ceiling the attacker has already lowered and maintaining the boundary
that prevents the attack from succeeding. ClawSentry maintains the boundary at a
moderate cost---6.7 points against that lowered ceiling---in exchange for
21.1--39.5 points of ASR reduction (Table~\ref{tab:skillsafety_main}) with clean
work preserved at 98.7\%; an operator who weights completion differently can
shift the operating point on measured terms (Appendix~\ref{app:fspr_modes}). The
margin should not, however, be read as protection paid for out of ordinary work.
The limit is architectural rather than parametric, which is why joint safety and
utility must be secured before the task begins: at the admission gate measured
in \S\ref{sec:exp:fspr}, and ultimately at distribution time
(\S\ref{sec:discussion}).

\begin{figure}[htbp]
\centering
\includegraphics[width=\linewidth]{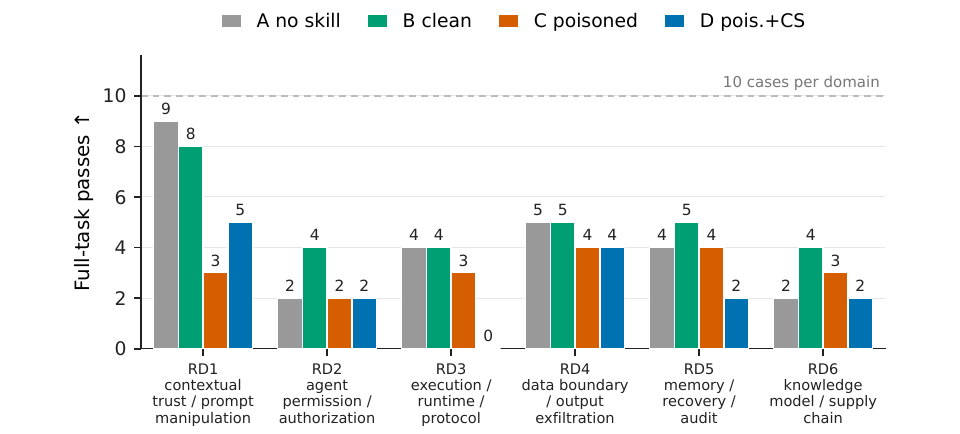}
\caption{The same four conditions broken down across the six risk domains, ten
cases each. The aggregate pattern is not carried by a single domain: a clean
skill raises the pass count above the no-skill baseline in RD2, RD5, and RD6,
whereas poisoning lowers it relative to the clean skill in all six domains.
RD1 shows the sharpest contrast, 8/10 with a clean skill versus 3/10 once the
same class of skill is poisoned, with ClawSentry disabled in both. The
enforcement margin from C to D is likewise concentrated rather than uniform.
Per-domain counts are tabulated in Appendix~\ref{app:tsr_analysis}.}
\label{fig:tsr_per_domain}
\end{figure}

\subsection{FSPR Package Discrimination}
\label{sec:exp:fspr}

\begin{figure}[htbp]
\centering
\includegraphics[width=\linewidth]{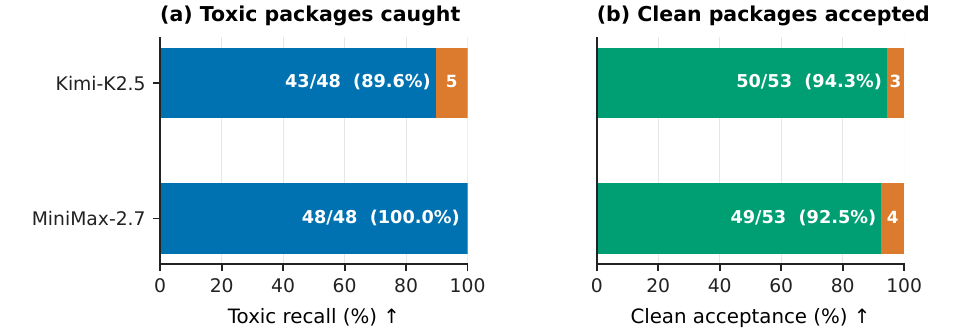}
\caption{FSPR discrimination on 48 toxic and 53 clean packages.
(a) Toxic-package recall for two diagnostic package-review backends.
(b) Clean-package acceptance under the same evaluation. Labels report exact
counts and percentages; orange segments denote missed toxic packages or
false-blocked clean packages.}
\label{fig:fspr_discrimination}
\end{figure}

End-to-end ASR combines package admission with later Work Agent behavior, so
we isolate FSPR on the 101-package mixed corpus
(Figure~\ref{fig:fspr_discrimination}). This study deliberately uses two review
backends other than the main-experiment reviewer, so that what it tests is
whether package-level discrimination is a property of the audit rather than of
one reviewer model: \texttt{minimax-m2.7} identifies all 48 toxic packages
while accepting 49 of 53 clean, and \texttt{kimi-k2.5} identifies 43 of 48
toxic while accepting 50 of 53 clean. Both combine high toxic recall with
clean over-blocking below 8\%, which is what makes package-level audit usable
as an admission gate rather than merely a detector, and the agreement of two
unrelated backends is what the comparison is for. Because neither backend is the
\texttt{gemini-3.5-flash} reviewer of the main runs, this 5.7--7.5\% residual
clean false-block rate estimates the scale of admission-time overblocking the
mechanism produces; it is not a direct measurement of the overblocking under
the main-reviewer configuration of \S\ref{sec:exp:utility} and
\S\ref{sec:exp:tsr_analysis}.

\subsection{Controlled Comparisons and System Analysis}
\label{sec:exp:ablation}

Table~\ref{tab:ablations} collects three controlled studies, each holding the
Work Agent, the 60-case fixture, and the reviewer fixed while varying one
factor; reviewer choice is treated separately below.

\begin{table}[htbp]
\centering
\small
\setlength{\tabcolsep}{4pt}
\begin{tabular}{@{}lcc@{}}
\toprule
\textbf{Study / condition} & \textbf{ASR}$\downarrow$ & \textbf{TSR}$\uparrow$ \\
\midrule
\multicolumn{3}{@{}l}{\emph{FSPR response to a flagged package} (CC/MiniMax-M3)} \\
\quad Hard block (default) & \textbf{8.47} & 18.64 \\
\quad Warning & 24.56 & 21.05 \\
\quad Runtime feedback & 10.53 & 17.54 \\
\midrule
\multicolumn{3}{@{}l}{\emph{Defense layer} (CC/MiniMax-M3)} \\
\quad ClawSentry & \textbf{8.62} & 18.97 \\
\quad \texttt{claude-guardrails}~\cite{dwarvesf2026guardrails} & 57.14 & 28.57 \\
\quad Raw (undefended) & 58.33 & 30.00 \\
\midrule
\multicolumn{3}{@{}l}{\emph{Component removal} (Codex/GPT-5.4)} \\
\quad Full & \textbf{12.73} & 25.45 \\
\quad $-$ FSPR & 23.73 & 30.51 \\
\quad $-$ Anti-bypass & 19.30 & 29.82 \\
\bottomrule
\end{tabular}
\caption{Controlled comparisons (\%). Each block is an independent study on its
own 60-case SkillsSafety subset with Work Agent, reviewer
(\texttt{gemini-3.5-flash}), and remaining settings fixed, so values compare
within a block, not across. Bold marks each block's lowest ASR;
Figure~\ref{fig:controlled_comparisons} plots each block against its TSR
operating point.}
\label{tab:ablations}
\end{table}

\begin{figure}[htbp]
\centering
\includegraphics[width=\linewidth]{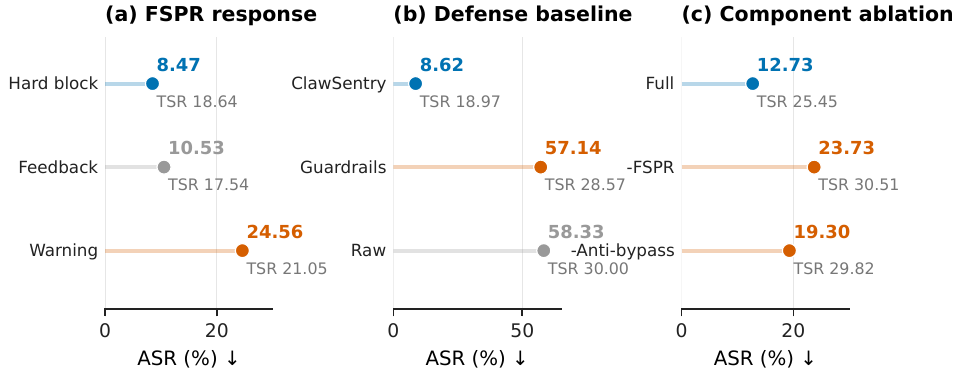}
\caption{Controlled security comparisons on fixed 60-case SkillsSafety
subsets. (a) FSPR response policy: hard block attains the lowest ASR of the
three responses, 2.06 points below runtime feedback and 16.09 points below a
document-level warning. (b) Matched defense baseline: ClawSentry reduces ASR by
48.52 points relative to Guardrails. (c) Component ablation: removing FSPR
or anti-bypass raises ASR by 11.00 and 6.57 points, respectively. Gray
annotations report the corresponding TSR operating points.}
\label{fig:controlled_comparisons}
\end{figure}

\paragraph{How should FSPR respond to a flagged package?}
Hard block defines the strongest observed operating point: it cuts ASR by
16.09 points relative to warning for a 2.41-point TSR difference, and
dominates runtime feedback on both axes. The gap is structural rather than
incidental: a soft response amounts to advice injected into a context the
attacker has already shaped. The user prompt may itself have requested the skill,
in which case it outranks a later caution, and by the time the warning is
delivered the model may already have read the package. Refusal at admission never
grants the package that channel, so ClawSentry retains a hard boundary
(Appendix~\ref{app:fspr_modes}).

\paragraph{Matched defense baseline.}
Against the open-source \texttt{claude-guardrails} PreToolUse
hooks~\cite{dwarvesf2026guardrails} and an
undefended configuration, with only the defense layer changing, ClawSentry
removes 48.52 and 49.71 points of ASR respectively. The rule-hook baseline
recovers almost none of that gap, because pattern hooks inspect an individual
tool invocation without a view of the package that produced it or of the
effects already denied in the session.

\paragraph{Component contributions.}
Removing FSPR costs 11.00 points of ASR and removing anti-bypass 6.57, so
package-level admission and capability-equivalent follow-up memory contribute
independently: admission review acts before any tool call exists, while
anti-bypass acts after a denial once the agent seeks an equivalent route.

\begin{figure}[htbp]
\centering
\includegraphics[width=\linewidth]{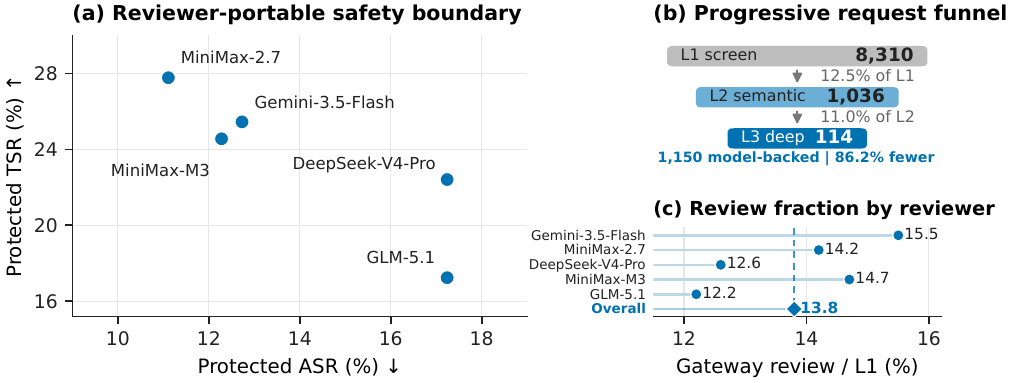}
\caption{Reviewer substitution and review volume, with Work Agent, policy, and
sample fixed. (a) Five reviewers share a compact protected ASR--TSR region.
(b) The aggregate progressive funnel across 300 provider--case executions: of
8{,}310 L1 events, 1{,}036 reach L2 and 114 reach L3. (c) Total Gateway review
stays at 12.2--15.5\% of L1 event volume across reviewers, compared with a
13.8\% aggregate.}
\label{fig:portability_funnel}
\end{figure}

\paragraph{Reviewer replaceability.}
With the Work Agent and sample fixed, substituting the provider used by Gateway
L2/L3 and FSPR keeps ASR within \textbf{11.11--17.24\%} across five providers
(Figure~\ref{fig:portability_funnel}(a)); \texttt{minimax-m2.7} attains both
the lowest ASR (11.11\%) and highest TSR (27.78\%). All remain far below the
33.5--49.7\% raw range, showing that the gateway-owned enforcement chain, not
one reviewer family, defines the security boundary
(\S\ref{sec:position}).

\subsection{Progressive Request Funnel}
\label{sec:exp:cost}

The case for progressive escalation concerns review volume rather than the cost
of any individual review. Across 300 provider--case executions
(Appendix~\ref{app:cost}), L1 processes 8{,}310 runtime events, 1{,}036 receive
L2 review, and only 114 issue a Gateway L3 request
(Figure~\ref{fig:portability_funnel}(b)). Counting L2 and L3 calls
separately, the Gateway makes 1{,}150 model-backed requests---\textbf{13.8\%}
of L1 event volume and \textbf{86.2\% fewer} than reviewing every event; the
ratio remains stable across providers (12.2--15.5\%,
Figure~\ref{fig:portability_funnel}(c)), and L3 alone is 1.37\% of L1 volume.
The funnel therefore confines heavyweight review to the genuinely
ambiguous residue, so the security results of \S\ref{sec:exp:security} come at
a review volume that scales with ambiguity rather than with tool-call traffic
(\S\ref{sec:position}).

\FloatBarrier

\section{Discussion: Why Admission-Time Review Dominates Runtime Intervention}
\label{sec:discussion}
\label{app:discussion:timing}

Two experiments in this paper were designed to answer different questions and
converge on one structural claim: \emph{an intervention whose first action comes
after a package has entered the agent's context operates from a disadvantaged
position, and the disadvantage is one no setting of its parameters removes.}

The utility diagnostic (\S\ref{sec:exp:tsr_analysis}) establishes the first
half.  Conditions A--C contain no ClawSentry component, so they measure the
attack side alone, and they show that poisoning removes 18.3 points of task
success relative to a clean same-class package---11.7 points below the agent's
own no-skill baseline---before any enforcement decision exists.  Whatever a
runtime guard does next, it operates on a package whose benign value has
already been destroyed: it can bound the damage, but it cannot restore the
utility, and the clean-skill reference is not a target a runtime mechanism can
be tuned toward.  Enforcement adds a further 6.7 points of overblocking
(\S\ref{sec:exp:tsr_analysis}), a margin we treat as the chosen price of the
boundary rather than recoverable headroom: recovering it means granting the
poisoned package the latitude the attack needs.

The response-mode study (Appendix~\ref{app:fspr_modes}) establishes the second
half from the enforcement side.  Neither soft response matches admission
refusal: a warning leaves ASR at 24.56\% and runtime feedback at 10.53\%,
against 8.47\% for refusal.  The three responses differ in how tightly they
couple to the decision itself, and ASR orders with that coupling, so the tightest
test is the one against runtime feedback---the strongest soft response
available, which still does not close on refusal.  The residue is positional
rather than parametric: feedback is advice delivered
into a context the attacker has already shaped, the user prompt may itself have
requested the skill and outranks a later caution, and by delivery time the model
may already have read the package.  Among the responses evaluated, admission
refusal alone never opens that channel.

Read together, the two results place both safety and utility \emph{before} the
task begins, and make the ordering architectural rather than a matter of tuning:
review before release, not attribution after the fact.  This is why FSPR rather
than the tiers is ClawSentry's first line of defense, and it locates the
remaining headroom outside the runtime gateway---in skill distribution channels
that carry admission evidence with the package.  Runtime supervision is the
necessary residual for what admission cannot settle: packages that are clean at
admission but steered later, and effects that only a session's accumulated
context reveals.

\section{Evaluation Scope and Measured Boundary}
\label{sec:limitations}

\paragraph{Benchmark-relative attacker scope.}
Both benchmarks were constructed independently of ClawSentry, so the reported
attacks are non-adaptive: no case was written with knowledge of the L1 short
circuits, the 25 shipped L2 patterns, L3 review manifests, or follow-up-guard
fingerprints. A policy-aware attacker can probe rewrites that clear a
deterministic floor or shape the evidence shown to L2/L3. The architecture
bounds the consequence: reviewer toolkits have no write, execution, or network
authority, so a subverted review turn costs evidence quality rather than host
state (\S\ref{sec:method:progressive}). The reported ASR therefore
characterizes the attack distributions encoded by SkillsSafety and
SkillInject; adaptive policy-aware evaluation is a distinct next setting.

\paragraph{Benchmark coverage.}
The evaluation spans two skill-facing benchmarks and two suites built from
their material. They cover third-party skill packages on coding-agent
harnesses; browser/GUI agents, long-horizon multi-agent delegation, and
non-English tasks define complementary deployment settings. With 155 and 319
cases, per-domain and per-condition cells support structural comparison rather
than fine-grained effect-size estimation.

\paragraph{Observed post-action consequences.}
Mainstream harnesses expose no locus-D hook that can retract or rewrite a
returned tool result, so post-action processing reports rather than enforces
(\S\ref{sec:method:post_action}). Its measurable contribution---whether
high-severity output evidence raises scrutiny on a subsequent action---depends
on whether asynchronous analysis completes before that action arrives.
Accordingly, the evaluation treats locus D as an evidence mechanism rather
than assigning it an isolated blocking number.

\paragraph{Provider-independent cost measure.}
\S\ref{sec:exp:cost} reports review volume, the architectural quantity the
gateway controls, rather than wall-clock or monetary cost dominated by
third-party queueing and rate limits. The measured result is that 86.2\% of L1
events close without a provider call. Latency and token accounting under
fixed, self-hosted reviewers can specialize that provider-independent bound
for a deployment.

\section{Conclusion}
\label{sec:conclusion}

Agentic skill risk is progressive across the lifecycle, and a denied effect
can re-enter across surface form, tool, or turn. ClawSentry counters this with a framework-agnostic
gateway combining pre-use package review, effect-normalized L1--L3 mediation,
anti-bypass memory, and post-action evidence, while keeping reviewers read-only
and policy authority gateway-owned. Across five Work Agents, it confines
SkillsSafety ASR to 9.09--15.03\% (33.5--49.7\% raw) at 98.7\% clean-skill TSR;
only 13.8\% of L1 events require model review. Its 6.7-point TSR cost locates
the enforcement boundary, while further gains lie in evidence-carrying
distribution channels and AHP-based multi-agent delegation.

\bibliography{paper/ref}

\begin{thebibliography}{53}
\providecommand{\natexlab}[1]{#1}
\providecommand{\url}[1]{\texttt{#1}}
\expandafter\ifx\csname urlstyle\endcsname\relax
  \providecommand{\doi}[1]{doi: #1}\else
  \providecommand{\doi}{doi: \begingroup \urlstyle{rm}\Url}\fi

\bibitem[Anthropic(2024)]{anthropic2024mcp}
Anthropic.
\newblock Model context protocol specification.
\newblock \url{https://modelcontextprotocol.io/specification}, 2024.

\bibitem[{Anthropic}(2025{\natexlab{a}})]{anthropic2025espionage}
{Anthropic}.
\newblock {Disrupting the First Reported AI-Orchestrated Cyber Espionage
  Campaign}.
\newblock \url{https://www.anthropic.com/news/disrupting-AI-espionage},
  November 2025{\natexlab{a}}.

\bibitem[{Anthropic}(2025{\natexlab{b}})]{anthropic2025threatintel}
{Anthropic}.
\newblock {Detecting and Countering Misuse of AI: August 2025}.
\newblock
  \url{https://www.anthropic.com/news/detecting-countering-misuse-aug-2025},
  August 2025{\natexlab{b}}.

\bibitem[{Anthropic}(2026{\natexlab{a}})]{anthropic2026mythos}
{Anthropic}.
\newblock {System Card: Claude Mythos Preview}.
\newblock
  \url{https://www-cdn.anthropic.com/08ab9158070959f88f296514c21b7facce6f52bc.pdf},
  April 2026{\natexlab{a}}.

\bibitem[{Anthropic}(2026{\natexlab{b}})]{anthropic2026mythosred}
{Anthropic}.
\newblock {Assessing Claude Mythos Preview's Cybersecurity Capabilities}.
\newblock \url{https://red.anthropic.com/2026/mythos-preview/}, April
  2026{\natexlab{b}}.

\bibitem[Beurer-Kellner et~al.(2025)Beurer-Kellner, Buesser, Creţu,
  Debenedetti, Dobos, Fabian, Fischer, Froelicher, Grosse, Naeff, Ozoani,
  Paverd, Tramèr, and Volhejn]{beurerkellner2025designpatterns}
Luca Beurer-Kellner, Beat Buesser, Ana-Maria Creţu, Edoardo Debenedetti,
  Daniel Dobos, Daniel Fabian, Marc Fischer, David Froelicher, Kathrin Grosse,
  Daniel Naeff, Ezinwanne Ozoani, Andrew Paverd, Florian Tramèr, and Václav
  Volhejn.
\newblock Design patterns for securing llm agents against prompt injections,
  2025.
\newblock URL \url{https://arxiv.org/abs/2506.08837}.

\bibitem[Chen et~al.(2026)Chen, Li, Zhou, Gong, Jiang, Dan, Zhang, Yu, Wang,
  Ma, Zhong, Zhu, Xiao, Yang, Du, Zhang, Zhang, Huang, Zhang, Du, Zhao, Guo,
  Chen, Ding, Sun, Li, Zhang, Yang, Yu, Zheng, Zheng, Li, Zhu, Zhou, Zhang,
  Ding, Zhang, Sun, Lyu, Lu, Wang, Shi, Li, Chen, Zhang, Zhuang, Cai, Pan, Li,
  Song, Zhang, Wang, Gu, Zhu, Dong, Li, Zhang, Zhuang, Tian, Liu, Hu, Tao,
  Zhang, Ruan, Xu, Yan, Liu, He, Xu, Ji, Yang, Xiao, Duan, Li, Han, Ruan, Yuan,
  Yu, Feng, Mo, Li, Du, Bao, Yang, Zhou, Loki, Chen, Zeng, Li, Zhong, Tao, Chi,
  Lin, Hu, Chen, Zhu, Gao, Gao, Li, Li, Zhao, Ren, Ren, Xu, Ren, Li, Wang,
  Chen, Zeng, Tian, Guo, Dong, Leng, Zhang, Liu, Chen, Chen, Jia, Yao, Zhao,
  Yu, Li, Pan, Zhu, Li, Xie, Qin, Li, Liang, Liu, Xu, Li, Chen, Cheng, Zhang,
  Chen, Zhao, Chen, Song, Wang, Luo, Luo, Su, Li, Han, Wu, Song, Han, Guan, Lu,
  Zou, Lai, Li, Shen, Gong, Ma, Jiao, Wang, Xu, Wang, Tang, Chen, Wang, Qiu,
  Shi, Guo, Huang, Wang, Hu, Gao, Zhang, Li, Ying, Zhang, Wang, Song, Yang,
  Meng, Miao, Li, Liu, Hu, Huang, Li, Huang, Zhang, Hong, Xie, Zhang, Liao,
  Shi, Wenren, Li, Li, Luo, Jin, Sun, Zhou, Su, Li, Zhu, Peng, Fan, Zhang, Xu,
  Lv, Xu, He, He, Li, Gao, Wu, Song, Zhou, Sun, Huang, Chen, and
  Ge]{chen2026minimaxm2seriesminiactivations}
Aili Chen, Aonian Li, Baichuan Zhou, Bangwei Gong, Binyang Jiang, Boji Dan,
  Changhao Zhang, Changqing Yu, Chao Wang, Cheng Ma, Cheng Zhong, Cheng Zhu,
  Chengjun Xiao, Chengyi Yang, Chengyu Du, Chenyang Zhang, Chi Zhang, Chuangyi
  Huang, Chunhao Zhang, Chunhui Du, Chunyu Zhao, Congchao Guo, Da~Chen, Deming
  Ding, Dianjun Sun, Dong Li, Dongyu Zhang, Enhui Yang, Fei Yu, Guang Zheng,
  Guodong Zheng, Guohong Li, Haichao Zhu, Haigang Zhou, Haimo Zhang, Han Ding,
  Hao Zhang, Haohai Sun, Haolin Lyu, Haonan Lu, Haoyu Wang, Huajie Shi, Huiyang
  Li, Jiacheng Chen, Jian Zhang, Jiaqi Zhuang, Jiaren Cai, Jiaxin Pan, Jiayao
  Li, Jiayuan Song, Jichuan Zhang, Jie Wang, Jihao Gu, Jin Zhu, Jingwei Dong,
  Jingyang Li, Jingyu Zhang, Jingze Zhuang, Jinhao Tian, Jinli Liu, Jinyi Hu,
  Jun Tao, Jun Zhang, Junbin Ruan, Junhao Xu, Junjie Yan, Junteng Liu, Junxian
  He, Kang Xu, Ke~Ji, Ke~Yang, Kecheng Xiao, Keyu Duan, Keyu Li, Le~Han, Letian
  Ruan, Li~Yuan, Lianfei Yu, Liheng Feng, Lijie Mo, Lin Li, Linge Du, Lingye
  Bao, Lingyu Yang, Lingyuan Zhou, Loki, Lu~Chen, Lunbin Zeng, Ming Li, Ming
  Zhong, Mingliang Tao, Mingyuan Chi, Mujie Lin, Nan Hu, Ningxin Chen, Peiyin
  Zhu, Peng Gao, Pengcheng Gao, Pengfei Li, Penglin Li, Pengyu Zhao, Qibin Ren,
  Qibing Ren, Qidi Xu, Qihan Ren, Qile Li, Qin Wang, Quanliang Chen, Qunhong
  Zeng, Rong Tian, Rongxin Guo, Rui Dong, Ruitao Leng, Ruize Zhang, Shanqi Liu,
  Shaoxiang Chen, Shaoyu Chen, Sheng Jia, Shun Yao, Shuoran Zhao, Shuqi Yu,
  Sichen Li, Sicheng Pan, Songquan Zhu, Tengfei Li, Tian Xie, Tiancheng Qin,
  Tianle Li, Tianrun Liang, Wei Liu, Weiqi Xu, Weitao Li, Weixiang Chen, Weiyu
  Cheng, Weiyu Zhang, Wenhu Chen, Wenqian Zhao, Xiancai Chen, Xiangjun Song,
  Xiangyuan Wang, Xianzhen Luo, Xiao Luo, Xiao Su, Xiaobo Li, Xiaodong Han,
  Xiaojie Wu, Xihao Song, Xingyi Han, Xinyu Guan, Xuan Lu, Xun Zou, Xunhao Lai,
  Xutong Li, Xuyang Shen, Yan Gong, Yan Ma, Yang Jiao, Yang Wang, Yang Xu,
  Yangsen Wang, Ye~Tang, Yicheng Chen, Yihang Wang, Yinran Qiu, Yiqi Shi,
  Yiting Guo, Yiwen Huang, Yixuan Wang, Yongyi Hu, Yu~Gao, Yu~Zhang, Yuan Li,
  Yuanxiang Ying, Yuanzhen Zhang, Yubo Wang, Yuchen Song, Yufeng Yang, Yuhang
  Meng, Yuhang Miao, Yuhao Li, Yujie Liu, Yulin Hu, Yunan Huang, Yunji Li,
  Yunyi Huang, Yusen Zhang, Yusu Hong, Yutao Xie, Yutong Zhang, Yuwen Liao,
  Yuxuan Shi, Yuze Wenren, Zebin Li, Zehan Li, Zejian Luo, Zeyu Jin, Zeyuan
  Sun, Zhanpeng Zhou, Zhaochen Su, Zhendong Li, Zhengmao Zhu, Zhengyuan Peng,
  Zhenhua Fan, Zhi Zhang, Zhichao Xu, Zhiheng Lv, Zhikang Xu, Zhitao He, Zhiwei
  He, Zhongyuan Li, Zibo Gao, Zijia Wu, Zijian Song, Zijian Zhou, Zijun Sun,
  Zishan Huang, Ziying Chen, and Ziyue Ge.
\newblock The minimax-m2 series: Mini activations unleashing max real-world
  intelligence, 2026.
\newblock URL \url{https://arxiv.org/abs/2605.26494}.

\bibitem[Debenedetti et~al.(2024)Debenedetti, Zhang, Balunovic, Beurer-Kellner,
  Fischer, and Tram{\`e}r]{debenedetti2024agentdojo}
Edoardo Debenedetti, Jie Zhang, Mislav Balunovic, Luca Beurer-Kellner, Marc
  Fischer, and Florian Tram{\`e}r.
\newblock Agentdojo: A dynamic environment to evaluate prompt injection attacks
  and defenses for {LLM} agents.
\newblock In \emph{The Thirty-eight Conference on Neural Information Processing
  Systems Datasets and Benchmarks Track}, 2024.
\newblock URL \url{https://openreview.net/forum?id=m1YYAQjO3w}.

\bibitem[Deng et~al.(2025)Deng, Guo, Han, Ma, Xiong, Wen, and
  Xiang]{deng2025ai}
Zehang Deng, Yongjian Guo, Changzhou Han, Wanlun Ma, Junwu Xiong, Sheng Wen,
  and Yang Xiang.
\newblock Ai agents under threat: A survey of key security challenges and
  future pathways.
\newblock \emph{ACM Comput. Surv.}, 57\penalty0 (7), February 2025.
\newblock ISSN 0360-0300.
\newblock \doi{10.1145/3716628}.
\newblock URL \url{https://doi.org/10.1145/3716628}.

\bibitem[{Dwarves Foundation}(2025)]{dwarvesf2026guardrails}
{Dwarves Foundation}.
\newblock claude-guardrails: Hardened security configuration for claude code.
\newblock \url{https://github.com/dwarvesf/claude-guardrails}, 2025.

\bibitem[Fang et~al.(2024)Fang, Bindu, Gupta, Zhan, and
  Kang]{fang2024llmagents}
Richard Fang, Rohan Bindu, Akul Gupta, Qiusi Zhan, and Daniel Kang.
\newblock Llm agents can autonomously hack websites, 2024.
\newblock URL \url{https://arxiv.org/abs/2402.06664}.

\bibitem[Ferrag et~al.(2026)Ferrag, Tihanyi, Hamouda, Maglaras, Lakas, and
  Debbah]{ferrag2025prompt}
Mohamed~Amine Ferrag, Norbert Tihanyi, Djallel Hamouda, Leandros Maglaras,
  Abderrahmane Lakas, and Merouane Debbah.
\newblock From prompt injections to protocol exploits: Threats in llm-powered
  ai agents workflows.
\newblock \emph{ICT Express}, 12\penalty0 (2):\penalty0 353--383, April 2026.
\newblock ISSN 2405-9595.
\newblock \doi{10.1016/j.icte.2025.12.001}.
\newblock URL \url{http://dx.doi.org/10.1016/j.icte.2025.12.001}.

\bibitem[Ghosh et~al.(2024)Ghosh, Varshney, Galinkin, and
  Parisien]{ghosh2024aegis}
Shaona Ghosh, Prasoon Varshney, Erick Galinkin, and Christopher Parisien.
\newblock Aegis: Online adaptive ai content safety moderation with ensemble of
  llm experts, 2024.
\newblock URL \url{https://arxiv.org/abs/2404.05993}.

\bibitem[{GLM-5-Team} et~al.(2026){GLM-5-Team}, Zeng, Lv, Hou, Du, Zheng, Chen,
  Yin, Ge, Huang, Xie, Zhu, Yin, Wang, Pan, Zeng, Zhang, Wang, Chen, Zhang,
  Jiao, Guo, Wang, Du, Wu, Wang, Li, Fan, Zhong, Liu, Zhao, Du, Dong, Lu,
  Shuang-Li, Cao, Liu, Jiang, Chen, Zhang, Huang, Dong, Xu, Wei, An, Niu, Zhu,
  Wen, Cen, Bai, Qiao, Wang, Wang, Zhu, Liu, Li, Wang, Wen, Huang, Cai, Yu, Li,
  Hu, Zhang, Zhang, Lin, Yang, Wang, Ai, Zhu, Yi, Chen, Wen, Sun, Zhao, Hu,
  Zhang, Liu, Zhang, Peng, Tai, Zhang, Liu, Wang, Yan, Ge, Liu, Chu, Zhao,
  Wang, Zhao, Ren, Wang, Zhang, Gui, Zhao, Li, An, Li, Yuan, Du, Liu, Zhi,
  Duan, Zhou, Wei, Wang, Luo, Zhang, Sha, Xu, Wu, Ding, Chen, Li, Lin, Ta, Zou,
  Song, Yang, Tu, Yang, Wu, Zhang, Li, Li, Fan, Qin, Tian, Zhang, Yu, Liang,
  Kuang, Cheng, Li, Yan, Hu, Ling, Fan, Xia, Zhang, Zhang, Pan, Zou, Zhang,
  Liu, Wu, Li, Wang, Zhu, Tan, Zhou, Pan, Zhang, Su, Geng, Yan, Tan, Bi, Shen,
  Yang, Li, Liu, Wang, Li, Wu, Zhang, Duan, Zhang, Liu, Jiang, Yan, Zhang, Wei,
  Chen, Feng, Yao, Chai, Wang, Zhang, Xu, Huang, Wang, Li, Dong, and
  Tang]{glm5team2026glm5vibecodingagentic}
{GLM-5-Team}, Aohan Zeng, Xin Lv, Zhenyu Hou, Zhengxiao Du, Qinkai Zheng, Bin
  Chen, Da~Yin, Chendi Ge, Chenghua Huang, Chengxing Xie, Chenzheng Zhu,
  Congfeng Yin, Cunxiang Wang, Gengzheng Pan, Hao Zeng, Haoke Zhang, Haoran
  Wang, Huilong Chen, Jiajie Zhang, Jian Jiao, Jiaqi Guo, Jingsen Wang,
  Jingzhao Du, Jinzhu Wu, Kedong Wang, Lei Li, Lin Fan, Lucen Zhong, Mingdao
  Liu, Mingming Zhao, Pengfan Du, Qian Dong, Rui Lu, Shuang-Li, Shulin Cao,
  Song Liu, Ting Jiang, Xiaodong Chen, Xiaohan Zhang, Xuancheng Huang, Xuezhen
  Dong, Yabo Xu, Yao Wei, Yifan An, Yilin Niu, Yitong Zhu, Yuanhao Wen, Yukuo
  Cen, Yushi Bai, Zhongpei Qiao, Zihan Wang, Zikang Wang, Zilin Zhu, Ziqiang
  Liu, Zixuan Li, Bojie Wang, Bosi Wen, Can Huang, Changpeng Cai, Chao Yu, Chen
  Li, Chengwei Hu, Chenhui Zhang, Dan Zhang, Daoyan Lin, Dayong Yang, Di~Wang,
  Ding Ai, Erle Zhu, Fangzhou Yi, Feiyu Chen, Guohong Wen, Hailong Sun, Haisha
  Zhao, Haiyi Hu, Hanchen Zhang, Hanrui Liu, Hanyu Zhang, Hao Peng, Hao Tai,
  Haobo Zhang, He~Liu, Hongwei Wang, Hongxi Yan, Hongyu Ge, Huan Liu, Huanpeng
  Chu, Jia'ni Zhao, Jiachen Wang, Jiajing Zhao, Jiamin Ren, Jiapeng Wang,
  Jiaxin Zhang, Jiayi Gui, Jiayue Zhao, Jijie Li, Jing An, Jing Li, Jingwei
  Yuan, Jinhua Du, Jinxin Liu, Junkai Zhi, Junwen Duan, Kaiyue Zhou, Kangjian
  Wei, Ke~Wang, Keyun Luo, Laiqiang Zhang, Leigang Sha, Liang Xu, Lindong Wu,
  Lintao Ding, Lu~Chen, Minghao Li, Nianyi Lin, Pan Ta, Qiang Zou, Rongjun
  Song, Ruiqi Yang, Shangqing Tu, Shangtong Yang, Shaoxiang Wu, Shengyan Zhang,
  Shijie Li, Shuang Li, Shuyi Fan, Wei Qin, Wei Tian, Weining Zhang, Wenbo Yu,
  Wenjie Liang, Xiang Kuang, Xiangmeng Cheng, Xiangyang Li, Xiaoquan Yan,
  Xiaowei Hu, Xiaoying Ling, Xing Fan, Xingye Xia, Xinyuan Zhang, Xinze Zhang,
  Xirui Pan, Xu~Zou, Xunkai Zhang, Yadi Liu, Yandong Wu, Yanfu Li, Yidong Wang,
  Yifan Zhu, Yijun Tan, Yilin Zhou, Yiming Pan, Ying Zhang, Yinpei Su, Yipeng
  Geng, Yong Yan, Yonglin Tan, Yuean Bi, Yuhan Shen, Yuhao Yang, Yujiang Li,
  Yunan Liu, Yunqing Wang, Yuntao Li, Yurong Wu, Yutao Zhang, Yuxi Duan, Yuxuan
  Zhang, Zezhen Liu, Zhengtao Jiang, Zhenhe Yan, Zheyu Zhang, Zhixiang Wei,
  Zhuo Chen, Zhuoer Feng, Zijun Yao, Ziwei Chai, Ziyuan Wang, Zuzhou Zhang, Bin
  Xu, Minlie Huang, Hongning Wang, Juanzi Li, Yuxiao Dong, and Jie Tang.
\newblock Glm-5: from vibe coding to agentic engineering, 2026.
\newblock URL \url{https://arxiv.org/abs/2602.15763}.

\bibitem[{Google DeepMind}(2026)]{google2026gemini35flash}
{Google DeepMind}.
\newblock {Gemini 3.5 Flash}.
\newblock \url{https://deepmind.google/models/model-cards/gemini-3-5-flash/},
  2026.

\bibitem[Greshake et~al.(2023)Greshake, Abdelnabi, Mishra, Endres, Holz, and
  Fritz]{greshake2023indirect}
Kai Greshake, Sahar Abdelnabi, Shailesh Mishra, Christoph Endres, Thorsten
  Holz, and Mario Fritz.
\newblock Not what you've signed up for: Compromising real-world llm-integrated
  applications with indirect prompt injection.
\newblock In \emph{Proceedings of the 16th ACM workshop on artificial
  intelligence and security}, pages 79--90, 2023.

\bibitem[Hines et~al.(2024)Hines, Lopez, Hall, Zarfati, Zunger, and
  Kiciman]{hines2024spotlighting}
Keegan Hines, Gary Lopez, Matthew Hall, Federico Zarfati, Yonatan Zunger, and
  Emre Kiciman.
\newblock Defending against indirect prompt injection attacks with
  spotlighting, 2024.
\newblock URL \url{https://arxiv.org/abs/2403.14720}.

\bibitem[Inan et~al.(2023)Inan, Upasani, Chi, Rungta, Iyer, Mao, Tontchev, Hu,
  Fuller, Testuggine, and Khabsa]{inan2023llamaguard}
Hakan Inan, Kartikeya Upasani, Jianfeng Chi, Rashi Rungta, Krithika Iyer,
  Yuning Mao, Michael Tontchev, Qing Hu, Brian Fuller, Davide Testuggine, and
  Madian Khabsa.
\newblock Llama guard: Llm-based input-output safeguard for human-ai
  conversations, 2023.
\newblock URL \url{https://arxiv.org/abs/2312.06674}.

\bibitem[Jin et~al.(2026)Jin, Wang, Wei, Wang, Zeng, Zhang, Yang, Qu, Hu, and
  Xu]{jin2026skillsafetybench}
Chang Jin, An~Wang, Zeming Wei, Kai Wang, Biaojie Zeng, Qiaosheng Zhang, Chao
  Yang, Jingjing Qu, Xia Hu, and Xingcheng Xu.
\newblock Skillsafetybench: Evaluating agent safety under skill-facing attack
  surfaces, 2026.
\newblock URL \url{https://arxiv.org/abs/2605.12015}.

\bibitem[{Kimi Team} et~al.(2026){Kimi Team}, Bai, Bai, Bao, Cai, Cao, Chai,
  Charles, Che, Chen, Chen, Chen, Chen, Chen, Chen, Chen, Chen, Chen, Chen,
  Chen, Chen, Chen, Chen, Chen, Chen, Chen, Chen, Chen, Chen, Cheng, Cheng,
  Chu, Cui, Deng, Diao, Ding, Dong, Dong, Dong, Dong, Du, Du, Du, Du, Du, Fan,
  Fang, Feng, Feng, Fu, Fu, Gao, Gao, Ge, Geng, Gong, Gong, Gongque, Gu, Gu,
  Gu, Guan, Guan, Guo, Hao, He, He, He, He, He, He, Hong, Hu, Hu, Hu, Hu,
  Huang, Huang, Huang, Huang, Jia, Jiang, Jiang, Jin, Jing, Lai, Li, Li, Li,
  Li, Li, Li, Li, Li, Li, Li, Li, Li, Li, Li, Li, Li, Li, Li, Li, Li, Li, Li,
  Li, Li, Liao, Lin, Lin, Lin, Lin, Lin, Liu, Liu, Liu, Liu, Liu, Liu, Liu,
  Liu, Liu, Liu, Liu, Liu, Liu, Liu, Liu, Liu, Liu, Lu, Lu, Lu, Luo, Luo, Luo,
  Luo, Ma, Mao, Mei, Men, Meng, Meng, Miao, Ni, Ouyang, Pan, Pang, Qian, Qin,
  Qin, Qiu, Qu, Shang, Shao, Shen, Shen, Shi, Shi, Shi, Song, Song, Song, Song,
  Su, Su, Su, Sui, Sun, Sun, Sun, Sung, Tai, Tang, Tang, Tang, Tang, Tao, Teng,
  Tian, Tian, Wang, Wang, Wang, Wang, Wang, Wang, Wang, Wang, Wang, Wang, Wang,
  Wang, Wang, Wang, Wang, Wang, Wang, Wang, Wang, Wang, Wang, Wang, Wang, Wang,
  Wang, Wang, Wang, Wang, Wang, Wang, Wang, Wang, Wang, Wang, Wang, Wang, Wang,
  Wang, Wang, Wei, Wei, Wen, Wen, Wu, Wu, Wu, Wu, Wu, Wu, Wu, Wu, Wu, Xiao,
  Xie, Xie, Xie, Xing, Xu, Xu, Xu, Xu, Xu, Xu, Xu, Xu, Xu, Xu, Xu, Xu, Xu, Xu,
  Xu, Yan, Yan, Yang, Yang, Yang, Yang, Yang, Yang, Yang, Yang, Yang, Yang,
  Yang, Yang, Yang, Yang, Yao, Ye, Ye, Ye, Ye, Yebo, Yin, Yu, Yu, Yu, Yu, Yuan,
  Yuan, Yuan, Yue, Zeng, Zha, Zhan, Zhang, Zhang, Zhang, Zhang, Zhang, Zhang,
  Zhang, Zhang, Zhang, Zhang, Zhang, Zhang, Zhang, Zhang, Zhang, Zhang, Zhang,
  Zhang, Zhang, Zhao, Zhao, Zhao, Zhao, Zhao, Zhao, Zhao, Zhao, Zheng, Zheng,
  Zheng, Zheng, Zhong, Zhong, Zhong, Zhou, Zhou, Zhou, Zhou, Zhu, Zhu, Zhu,
  Zhu, Zhu, Zhuang, Zhuang, Zou, and Zu]{kimiteam2026kimik25visualagentic}
{Kimi Team}, Tongtong Bai, Yifan Bai, Yiping Bao, S.~H. Cai, Yuan Cao, Ziwei
  Chai, Y.~Charles, H.~S. Che, Cheng Chen, Guanduo Chen, Huarong Chen, Jia
  Chen, Jianlong Chen, Jun Chen, Kefan Chen, Liang Chen, Ruijue Chen, Xinhao
  Chen, Yanru Chen, Yanxu Chen, Yicun Chen, Yimin Chen, Yingjiang Chen, Yuankun
  Chen, Yujie Chen, Yutian Chen, Zhirong Chen, Ziwei Chen, Dazhi Cheng, Yean
  Cheng, Minghan Chu, Jialei Cui, Jiaqi Deng, Muxi Diao, Hao Ding, Mengfan
  Dong, Mengnan Dong, Yuxin Dong, Yuhao Dong, Angang Du, Chenzhuang Du, Dikang
  Du, Lingxiao Du, Yulun Du, Yu~Fan, Shengjun Fang, Qiulin Feng, Yichen Feng,
  Garimugai Fu, Kelin Fu, Hongcheng Gao, Tong Gao, Yuyao Ge, Shangyi Geng,
  Chengyang Gong, Xiaochen Gong, Zhuoma Gongque, Qizheng Gu, Xinran Gu, Yicheng
  Gu, Longyu Guan, Shuhao Guan, Yuanying Guo, Xiaoru Hao, Dailan He, Tianhong
  He, Weiran He, Wenyang He, Yibo He, Yunjia He, Chao Hong, Hao Hu, Jiaxi Hu,
  Yangyang Hu, Zhenxing Hu, Ke~Huang, Ruiyuan Huang, Weixiao Huang, Zhiqi
  Huang, Chaobo Jia, Tao Jiang, Zhejun Jiang, Xinyi Jin, Yu~Jing, Guokun Lai,
  Aidi Li, C.~Li, Cheng Li, Fang Li, Guanghe Li, Guanyu Li, Haitao Li, Haoyang
  Li, Jia Li, Jingwei Li, Junxiong Li, Lincan Li, Mo~Li, Weihong Li, Wentao Li,
  Xinhang Li, Xinhao Li, Yang Li, Yanhao Li, Yiwei Li, Yuxiao Li, Zhaowei Li,
  Zhaoxi Li, Zheming Li, Weilong Liao, Jiawei Lin, Xiaohan Lin, Yibo Lin,
  Zhishan Lin, Zichao Lin, Cheng Liu, Chenyu Liu, Hongzhang Liu, Liang Liu,
  Shaowei Liu, Shudong Liu, Shuran Liu, Tianwei Liu, Tianyu Liu, Weizhou Liu,
  Xiangyan Liu, Yangyang Liu, Yanming Liu, Yibo Liu, Yuanxin Liu, Zhengying
  Liu, Zhongnuo Liu, Enzhe Lu, Haoyu Lu, Zhiyuan Lu, G.~Luo, Junyu Luo, Tongxu
  Luo, Yashuo Luo, Long Ma, Shaoguang Mao, Yuan Mei, Xin Men, Fanqing Meng,
  Zhiyong Meng, Yibo Miao, Minqing Ni, Kun Ouyang, Siyuan Pan, Bo~Pang, Yuchao
  Qian, Ruoyu Qin, Zeyu Qin, Jiezhong Qiu, Bowen Qu, Zeyu Shang, Youbo Shao,
  Tianxiao Shen, Zhennan Shen, Juanfeng Shi, Lidong Shi, Shengyuan Shi, Feifan
  Song, Pengwei Song, Tianhui Song, Xiaoxi Song, Hongjin Su, Jianlin Su,
  Zhaochen Su, Lin Sui, Jinsong Sun, Junyao Sun, Tongyu Sun, Flood Sung,
  Yunpeng Tai, Chuning Tang, Heyi Tang, Xiaojuan Tang, Zhengyang Tang, Jiawen
  Tao, Shiyuan Teng, Chaoran Tian, Pengfei Tian, Bowen Wang, Chensi Wang,
  Chuang Wang, Congcong Wang, Dingkun Wang, Dinglu Wang, Dongliang Wang, Feng
  Wang, Hailong Wang, Haiming Wang, Hao Wang, Hengzhi Wang, Huaqing Wang, Hui
  Wang, Jiahao Wang, Jinhong Wang, Jiuzheng Wang, Kaixin Wang, Linian Wang,
  Qibin Wang, Shengjie Wang, Shuyi Wang, Si~Wang, Wei Wang, Xiaochen Wang,
  Xinyuan Wang, Yao Wang, Yejie Wang, Yipu Wang, Yiqin Wang, Yucheng Wang,
  Yuzhi Wang, Zhaoji Wang, Zhaowei Wang, Zhengtao Wang, Zhexu Wang, Zifan Wang,
  Zihan Wang, Zizhe Wang, Chu Wei, Ming Wei, Chuan Wen, Zichen Wen, Chengjie
  Wu, Haoning Wu, Junyan Wu, Rucong Wu, Wenhao Wu, Yuefeng Wu, Yuhao Wu, Yuxin
  Wu, Zijian Wu, Chenjun Xiao, Jin Xie, Xiaotong Xie, Yuchong Xie, Bowei Xing,
  Boyu Xu, Jianfan Xu, Jing Xu, Jinjing Xu, L.~H. Xu, Lin Xu, Suting Xu, Weixin
  Xu, Xinbo Xu, Xinran Xu, Yangchuan Xu, Yichang Xu, Yuemeng Xu, Zelai Xu,
  Ziyao Xu, Junjie Yan, Yuzi Yan, Guangyao Yang, Hao Yang, Junwei Yang, Kai
  Yang, Ningyuan Yang, Xiaofei Yang, Xinlong Yang, Xinyu Yang, Ying Yang,
  Yi~Yang, Yi~Yang, Zhen Yang, Zhilin Yang, Zonghan Yang, Haotian Yao, Dan Ye,
  Haoran Ye, Wenjie Ye, Zhuorui Ye, Peng Yebo, Bohong Yin, Chengzhen Yu,
  Longhui Yu, Tao Yu, Tianxiang Yu, Enming Yuan, Mengjie Yuan, Xiaokun Yuan,
  Yang Yue, Weihao Zeng, Dunyuan Zha, Haobing Zhan, Dehao Zhang, Hao Zhang, Jin
  Zhang, Puqi Zhang, Qiao Zhang, Rui Zhang, Xiaobin Zhang, Xiaoyun Zhang,
  Y.~Zhang, Yadong Zhang, Yangkun Zhang, Yichi Zhang, Yizhi Zhang, Yongting
  Zhang, Yu~Zhang, Yushun Zhang, Yutao Zhang, Yutong Zhang, Zheng Zhang,
  Chenguang Zhao, Feifan Zhao, Jinxiang Zhao, Shuai Zhao, Xiangyu Zhao, Xuanle
  Zhao, Yikai Zhao, Zijia Zhao, Huabin Zheng, Ruihan Zheng, Shaojie Zheng,
  Tengyang Zheng, Junfeng Zhong, Longguang Zhong, Weiming Zhong, M.~Zhou,
  Runjie Zhou, Xinyu Zhou, Zaida Zhou, Jinguo Zhu, Liya Zhu, Xinhao Zhu, Yuxuan
  Zhu, Zhen Zhu, Jingze Zhuang, Weiyu Zhuang, Ying Zou, and Xinxing Zu.
\newblock Kimi k2.5: Visual agentic intelligence, 2026.
\newblock URL \url{https://arxiv.org/abs/2602.02276}.

\bibitem[Kong et~al.(2025)Kong, Lin, Xu, Wang, Li, Li, Zhang, Peng, Chen, Sha,
  Li, Lin, Wang, Liu, Zhang, Chen, Wu, Khan, and Han]{kong2025survey}
Dezhang Kong, Shi Lin, Zhenhua Xu, Zhebo Wang, Minghao Li, Yufeng Li, Yilun
  Zhang, Hujin Peng, Xiang Chen, Zeyang Sha, Yuyuan Li, Changting Lin, Xun
  Wang, Xuan Liu, Ningyu Zhang, Chaochao Chen, Chunming Wu, Muhammad~Khurram
  Khan, and Meng Han.
\newblock A survey of llm-driven ai agent communication: Protocols, security
  risks, and defense countermeasures, 2025.
\newblock URL \url{https://arxiv.org/abs/2506.19676}.

\bibitem[Lai et~al.(2026)Lai, Xu, Yang, Chen, Xu, Zeng, Li, Sun, Zhu, Zhang,
  Hu, Li, Gao, Li, Zhu, Zhou, and Zhao]{lai2026minimaxsparseattention}
Xunhao Lai, Weiqi Xu, Yufeng Yang, Qiaorui Chen, Yang Xu, Lunbin Zeng, Xiaolong
  Li, Haohai Sun, Haichao Zhu, Vito Zhang, Jinkai Hu, Jiayao Li, Rui Gao, Zekun
  Li, Songquan Zhu, Jingkai Zhou, and Pengyu Zhao.
\newblock Minimax sparse attention, 2026.
\newblock URL \url{https://arxiv.org/abs/2606.13392}.

\bibitem[Li et~al.(2025)Li, Liu, Chun, Li, Zhang, and Xiao]{li2025drift}
Hao Li, Xiaogeng Liu, CHIU~Hung Chun, Dianqi Li, Ning Zhang, and Chaowei Xiao.
\newblock {DRIFT}: Dynamic rule-based defense with injection isolation for
  securing {LLM} agents.
\newblock In \emph{The Thirty-ninth Annual Conference on Neural Information
  Processing Systems}, 2025.
\newblock URL \url{https://openreview.net/forum?id=oY1Xnt83oJ}.

\bibitem[Li et~al.(2026)Li, Liu, Chen, You, Di, He, Zheng, Choe, Sun, Wang,
  Tao, Li, Zhao, Geng, Wu, Zhou, Chen, Xing, Li, Zeng, Wang, Wang, Chaim,
  Jiang, Shen, Kong, Liu, Wang, Liu, Li, Lan, Lin, Ye, He, Li, Zhang, Gao, Li,
  Ma, Jing, Wang, Li, Xue, Lyu, He, Tian, Wu, Wang, Gao, Chen, Liu, Cheng, Bao,
  Tong, Xu, Zhuo, Ye, Qi, Li, Liao, Tan, Shi, Tang, Tankasala, Yuan, Qian, Tu,
  Wang, Sun, Wang, Taylor, Yang, Guan, Dong, Zhang, Dillmann, chung Lee, and
  Song]{li2026skillsbench}
Xiangyi Li, Yimin Liu, Wenbo Chen, Bingran You, Zonglin Di, Yifeng He, Shenghan
  Zheng, Kyoung~Whan Choe, Jiankai Sun, Shuyi Wang, Chujun Tao, Binxu Li,
  Xuandong Zhao, Hejia Geng, Xiaojun Wu, Junwei Zhou, Xiaokun Chen, Hanwen
  Xing, Yubo Li, Qunhong Zeng, Di~Wang, Yuanli Wang, Roey~Ben Chaim, Penghao
  Jiang, Haotian Shen, Luyang Kong, Xinyi Liu, Runhui Wang, Xuanqing Liu,
  Jiachen Li, Xin Lan, Yueqian Lin, Wengao Ye, Junwei He, Songlin Li, Yue
  Zhang, Yipeng Gao, Yijiang Li, Ze~Ma, Liqiang Jing, Tianyu Wang, Kaixin Li,
  Yiqi Xue, Haoran Lyu, Yizhuo He, Yuchen Tian, Shutong Wu, Bowei Wang, Yixuan
  Gao, Bo~Chen, Litong Liu, Sikai Cheng, Jiajun Bao, Shuaicheng Tong, Shuwen
  Xu, Terry~Yue Zhuo, Tinghan Ye, Qi~Qi, Miao Li, Longtai Liao, Zelin Tan,
  Chang Shi, Xilin Tang, Srinath Tankasala, Boqin Yuan, Yaoyao Qian, Jianhong
  Tu, Chenguang Wang, Yizhou Sun, Wei Wang, Aaron Taylor, Ziyue Yang, Changkun
  Guan, Zhikang Dong, Xinyu Zhang, Steven Dillmann, Han chung Lee, and Dawn
  Song.
\newblock Skillsbench: Benchmarking how well agent skills work across diverse
  tasks, 2026.
\newblock URL \url{https://arxiv.org/abs/2602.12670}.

\bibitem[Liu et~al.(2026)Liu, Ren, Qian, Shao, Xie, Li, Yang, Luo, Wang, Liu,
  Hu, Tang, Mei, Guo, Yuan, Yang, Chen, Lin, Yu, Zhang, Guo, Zhang, Shao, Deng,
  Xi, Wang, Wang, Shen, Chen, Xie, Tao, Dai, Ji, Ba, Zhang, Liu, Zhang, Zhu,
  Wei, Xue, Lu, Shao, and Hu]{liu2026agentdog}
Dongrui Liu, Qihan Ren, Chen Qian, Shuai Shao, Yuejin Xie, Yu~Li, Zhonghao
  Yang, Haoyu Luo, Peng Wang, Qingyu Liu, Binxin Hu, Ling Tang, Jilin Mei, Dadi
  Guo, Leitao Yuan, Junyao Yang, Guanxu Chen, Qihao Lin, Yi~Yu, Bo~Zhang,
  Jiaxuan Guo, Jie Zhang, Wenqi Shao, Huiqi Deng, Zhiheng Xi, Wenjie Wang,
  Wenxuan Wang, Wen Shen, Zhikai Chen, Haoyu Xie, Jialing Tao, Juntao Dai,
  Jiaming Ji, Zhongjie Ba, Linfeng Zhang, Yong Liu, Quanshi Zhang, Lei Zhu,
  Zhihua Wei, Hui Xue, Chaochao Lu, Jing Shao, and Xia Hu.
\newblock Agentdog: A diagnostic guardrail framework for ai agent safety and
  security, 2026.
\newblock URL \url{https://arxiv.org/abs/2601.18491}.

\bibitem[Liu et~al.(2025)Liu, Deng, Li, Wang, Wang, Wang, Zhang, Liu, Wang,
  Zheng, Zhang, and Liu]{liu2023prompt}
Yi~Liu, Gelei Deng, Yuekang Li, Kailong Wang, Zihao Wang, Xiaofeng Wang,
  Tianwei Zhang, Yepang Liu, Haoyu Wang, Yan Zheng, Leo~Yu Zhang, and Yang Liu.
\newblock Prompt injection attack against llm-integrated applications, 2025.
\newblock URL \url{https://arxiv.org/abs/2306.05499}.

\bibitem[Lupinacci et~al.(2026)Lupinacci, Pironti, Blefari, Romeo, Arena, and
  Furfaro]{lupinacci2025dark}
Matteo Lupinacci, Francesco~Aurelio Pironti, Francesco Blefari, Francesco
  Romeo, Luigi Arena, and Angelo Furfaro.
\newblock \emph{The Dark Side of LLMs: Agent-based Attack Vectors
  for System-level Compromise}, pages 5--23.
\newblock Springer Nature Switzerland, August 2026.
\newblock ISBN 9783032355867.
\newblock \doi{10.1007/978-3-032-35586-7_1}.
\newblock URL \url{http://dx.doi.org/10.1007/978-3-032-35586-7_1}.

\bibitem[Ma et~al.(2025)Ma, Gao, Wang, Wang, Wang, Sun, Ding, Xu, Chen, Zhao,
  Huang, Li, Wu, Zhang, Zheng, Bai, Li, Wu, Qiu, Zhang, Han, Li, Sun, Wang, Gu,
  Wu, Chen, Zhang, Liu, Gong, Liu, Pan, Xie, Pang, Dong, Jia, Zhang, Ma, Zhang,
  Gong, Xiao, Erfani, Baldwin, Li, Sugiyama, Tao, Bailey, and
  Jiang]{ma2026safety}
Xingjun Ma, Yifeng Gao, Yixu Wang, Ruofan Wang, Xin Wang, Ye~Sun, Yifan Ding,
  Hengyuan Xu, Yunhao Chen, Yunhan Zhao, Hanxun Huang, Yige Li, Yutao Wu,
  Jiaming Zhang, Xiang Zheng, Yang Bai, Yiming Li, Zuxuan Wu, Xipeng Qiu,
  Jingfeng Zhang, Xudong Han, Haonan Li, Jun Sun, Cong Wang, Jindong Gu,
  Baoyuan Wu, Siheng Chen, Tianwei Zhang, Yang Liu, Mingming Gong, Tongliang
  Liu, Shirui Pan, Cihang Xie, Tianyu Pang, Yinpeng Dong, Ruoxi Jia, Yang
  Zhang, Shiqing Ma, Xiangyu Zhang, Neil Gong, Chaowei Xiao, Sarah Erfani, Tim
  Baldwin, Bo~Li, Masashi Sugiyama, Dacheng Tao, James Bailey, and Yu-Gang
  Jiang.
\newblock Safety at scale: a comprehensive survey of large model and agent
  safety.
\newblock \emph{Foundations and Trends in Privacy and Security}, 8\penalty0
  (3-4):\penalty0 1--240, 09 2025.
\newblock ISSN 2474-1558.
\newblock \doi{10.1561/3300000051}.
\newblock URL \url{https://doi.org/10.1561/3300000051}.

\bibitem[{Microsoft Security}(2026)]{microsoft2026mdash}
{Microsoft Security}.
\newblock {Defense at AI speed: Microsoft's new multi-model agentic security
  system tops leading industry benchmark}.
\newblock
  \url{https://www.microsoft.com/en-us/security/blog/2026/05/12/defense-at-ai-speed-microsofts-new-multi-model-agentic-security-system-tops-leading-industry-benchmark/},
  May 2026.

\bibitem[Naik et~al.(2026)Naik, Naik, and Naik]{naik2025insecure}
Dishita Naik, Ishita Naik, and Nitin Naik.
\newblock Insecure output handling in large language models (llms) and
  approaches to enhance output security, including prevention of llm-based web
  application attacks.
\newblock In Nitin Naik, Paul Jenkins, Shaligram Prajapat, and Paul Grace,
  editors, \emph{Contributions Presented at the International Conference on
  Computing, Communication, Cybersecurity {\&} AI, July 10--11, 2025,
  Birmingham, UK}, pages 695--720, Cham, 2026. Springer Nature Switzerland.
\newblock ISBN 978-3-032-16791-0.

\bibitem[{OpenAI}(2026{\natexlab{a}})]{openai2026disrupting}
{OpenAI}.
\newblock {Disrupting Malicious Uses of AI}.
\newblock \url{https://openai.com/index/disrupting-malicious-ai-uses/},
  February 2026{\natexlab{a}}.

\bibitem[{OpenAI}(2026{\natexlab{b}})]{openai2026gpt54}
{OpenAI}.
\newblock {GPT-5.4}.
\newblock \url{https://developers.openai.com/api/docs/models/gpt-5.4},
  2026{\natexlab{b}}.

\bibitem[{OpenAI}(2026{\natexlab{c}})]{openai2026gpt55}
{OpenAI}.
\newblock {GPT-5.5}.
\newblock \url{https://developers.openai.com/api/docs/models/gpt-5.5},
  2026{\natexlab{c}}.

\bibitem[{OWASP GenAI Security Project}(2025)]{owasp_agentic_2026}
{OWASP GenAI Security Project}.
\newblock {OWASP Top 10 for Agentic Applications 2026}.
\newblock Whitepaper, OWASP Foundation, 12 2025.
\newblock URL
  \url{https://genai.owasp.org/resource/owasp-top-10-for-agentic-applications-for-2026/}.

\bibitem[{Protect AI}(2023)]{protectai2023rebuff}
{Protect AI}.
\newblock Rebuff: Llm prompt injection detector.
\newblock GitHub repository, \url{https://github.com/protectai/rebuff}, 2023.

\bibitem[Rebedea et~al.(2023)Rebedea, Dinu, Sreedhar, Parisien, and
  Cohen]{rebedea2023nemo}
Traian Rebedea, Razvan Dinu, Makesh~Narsimhan Sreedhar, Christopher Parisien,
  and Jonathan Cohen.
\newblock {N}e{M}o guardrails: A toolkit for controllable and safe {LLM}
  applications with programmable rails.
\newblock In Yansong Feng and Els Lefever, editors, \emph{Proceedings of the
  2023 Conference on Empirical Methods in Natural Language Processing: System
  Demonstrations}, pages 431--445, Singapore, December 2023. Association for
  Computational Linguistics.
\newblock \doi{10.18653/v1/2023.emnlp-demo.40}.
\newblock URL \url{https://aclanthology.org/2023.emnlp-demo.40/}.

\bibitem[Ruan et~al.(2024)Ruan, Dong, Wang, Pitis, Zhou, Ba, Dubois, Maddison,
  and Hashimoto]{ruan2024toolemu}
Yangjun Ruan, Honghua Dong, Andrew Wang, Silviu Pitis, Yongchao Zhou, Jimmy Ba,
  Yann Dubois, Chris~J. Maddison, and Tatsunori Hashimoto.
\newblock Identifying the risks of {LM} agents with an {LM}-emulated sandbox.
\newblock In \emph{The Twelfth International Conference on Learning
  Representations}, 2024.
\newblock URL \url{https://openreview.net/forum?id=GEcwtMk1uA}.

\bibitem[Saltzer and Schroeder(1975)]{saltzer1975protection}
Jerome~H Saltzer and Michael~D Schroeder.
\newblock The protection of information in computer systems.
\newblock \emph{Proceedings of the IEEE}, 63\penalty0 (9):\penalty0 1278--1308,
  1975.

\bibitem[Sandhu et~al.(1996)Sandhu, Coyne, Feinstein, and
  Youman]{sandhu1996rbac}
Ravi~S Sandhu, Edward~J Coyne, Hal~L Feinstein, and Charles~E Youman.
\newblock Role-based access control models.
\newblock \emph{Computer}, 29\penalty0 (2):\penalty0 38--47, 1996.

\bibitem[Schmotz et~al.(2026)Schmotz, Beurer-Kellner, Abdelnabi, and
  Andriushchenko]{schmotz2026skillinject}
David Schmotz, Luca Beurer-Kellner, Sahar Abdelnabi, and Maksym Andriushchenko.
\newblock Skill-inject: Measuring agent vulnerability to skill file attacks,
  2026.
\newblock URL \url{https://arxiv.org/abs/2602.20156}.

\bibitem[{Significant Gravitas}(2023)]{autogpt2023}
{Significant Gravitas}.
\newblock Autogpt: An autonomous {GPT-4} experiment.
\newblock \url{https://github.com/Significant-Gravitas/AutoGPT}, 2023.

\bibitem[Wang et~al.(2025{\natexlab{a}})Wang, Poskitt, and
  Sun]{wang2025agentspec}
Haoyu Wang, Christopher~M. Poskitt, and Jun Sun.
\newblock Agentspec: Customizable runtime enforcement for safe and reliable llm
  agents, 2025{\natexlab{a}}.
\newblock URL \url{https://arxiv.org/abs/2503.18666}.

\bibitem[Wang et~al.(2025{\natexlab{b}})Wang, Li, Song, Xu, Tang, Zhuge, Pan,
  Song, Li, Singh, et~al.]{wang2025openhands}
Xingyao Wang, Boxuan Li, Yufan Song, Frank~F Xu, Xiangru Tang, Mingchen Zhuge,
  Jiayi Pan, Yueqi Song, Bowen Li, Jaskirat Singh, et~al.
\newblock Openhands: An open platform for ai software developers as generalist
  agents.
\newblock In \emph{International Conference on Learning Representations},
  volume 2025, pages 65882--65919, 2025{\natexlab{b}}.

\bibitem[Wang et~al.(2026)Wang, Shi, He, Cai, Zhang, and
  Song]{wang2026cybergym}
Zhun Wang, Tianneng Shi, Jingxuan He, Matthew Cai, Jialin Zhang, and Dawn Song.
\newblock Cybergym: Evaluating ai agents' real-world cybersecurity capabilities
  at scale.
\newblock In \emph{International Conference on Learning Representations},
  volume 2026, pages 123341--123386, 2026.

\bibitem[Wu et~al.(2024)Wu, Bansal, Zhang, Wu, Li, Zhu, Jiang, Zhang, Zhang,
  Liu, Awadallah, White, Burger, and Wang]{wu2024autogen}
Qingyun Wu, Gagan Bansal, Jieyu Zhang, Yiran Wu, Beibin Li, Erkang Zhu,
  Li~Jiang, Xiaoyun Zhang, Shaokun Zhang, Jiale Liu, Ahmed~Hassan Awadallah,
  Ryen~W. White, Doug Burger, and Chi Wang.
\newblock {AutoGen}: Enabling next-gen {LLM} applications via multi-agent
  conversations.
\newblock In \emph{First Conference on Language Modeling}, 2024.
\newblock URL \url{https://openreview.net/forum?id=BAakY1hNKS}.

\bibitem[Wu et~al.(2025)Wu, Roesner, Kohno, Zhang, and Iqbal]{wu2025isolategpt}
Yuhao Wu, Franziska Roesner, Tadayoshi Kohno, Ning Zhang, and Umar Iqbal.
\newblock Isolategpt: An execution isolation architecture for llm-based agentic
  systems.
\newblock In \emph{32nd Annual Network and Distributed System Security
  Symposium, {NDSS} 2025, San Diego, California, USA, February 24-28, 2025}.
  The Internet Society, 2025.
\newblock URL
  \url{https://www.ndss-symposium.org/ndss-paper/isolategpt-an-execution-isolation-architecture-for-llm-based-agentic-systems/}.

\bibitem[Yan et~al.(2025)Yan, Zhou, Zhang, Li, Zeng, Qi, Wang, and
  Zhang]{yan2025attack}
Bingyu Yan, Ziyi Zhou, Xiaoming Zhang, Chaozhuo Li, Ruilin Zeng, Yirui Qi,
  Tianbo Wang, and Litian Zhang.
\newblock Attack the messages, not the agents: A multi-round adaptive stealthy
  tampering framework for llm-mas, 2025.
\newblock URL \url{https://arxiv.org/abs/2508.03125}.

\bibitem[Yan et~al.(2026)Yan, Zhou, Zhang, Zhang, Zhou, Miao, Li, Li, and
  Zhang]{yan2025beyond}
Bingyu Yan, Zhibo Zhou, Litian Zhang, Lian Zhang, Ziyi Zhou, Dezhuang Miao,
  Zhoujun Li, Chaozhuo Li, and Xiaoming Zhang.
\newblock Beyond self-talk: A communication-centric survey of llm-based
  multi-agent systems, 2026.
\newblock URL \url{https://arxiv.org/abs/2502.14321}.

\bibitem[Yao et~al.(2023)Yao, Zhao, Yu, Du, Shafran, Narasimhan, and
  Cao]{yao2023react}
Shunyu Yao, Jeffrey Zhao, Dian Yu, Nan Du, Izhak Shafran, Karthik~R Narasimhan,
  and Yuan Cao.
\newblock React: Synergizing reasoning and acting in language models.
\newblock In \emph{The Eleventh International Conference on Learning
  Representations}, 2023.
\newblock URL \url{https://openreview.net/forum?id=WE_vluYUL-X}.

\bibitem[Yi et~al.(2025)Yi, Xie, Zhu, Kiciman, Sun, Xie, and Wu]{yi2023bipia}
Jingwei Yi, Yueqi Xie, Bin Zhu, Emre Kiciman, Guangzhong Sun, Xing Xie, and
  Fangzhao Wu.
\newblock Benchmarking and defending against indirect prompt injection attacks
  on large language models.
\newblock In \emph{Proceedings of the 31st ACM SIGKDD Conference on Knowledge
  Discovery and Data Mining V.1}, KDD ’25, pages 1809--1820. ACM, July 2025.
\newblock \doi{10.1145/3690624.3709179}.
\newblock URL \url{http://dx.doi.org/10.1145/3690624.3709179}.

\bibitem[Zeng et~al.(2024)Zeng, Liu, Mullins, Peran, Fernandez, Harkous,
  Narasimhan, Proud, Kumar, Radharapu, Sturman, and Wahltinez]{shieldgemma2024}
Wenjun Zeng, Yuchi Liu, Ryan Mullins, Ludovic Peran, Joe Fernandez, Hamza
  Harkous, Karthik Narasimhan, Drew Proud, Piyush Kumar, Bhaktipriya Radharapu,
  Olivia Sturman, and Oscar Wahltinez.
\newblock Shieldgemma: Generative ai content moderation based on gemma, 2024.
\newblock URL \url{https://arxiv.org/abs/2407.21772}.

\bibitem[Zhan et~al.(2024)Zhan, Liang, Ying, and Kang]{zhan2024injecagent}
Qiusi Zhan, Zhixiang Liang, Zifan Ying, and Daniel Kang.
\newblock Injecagent: Benchmarking indirect prompt injections in
  tool-integrated large language model agents.
\newblock In \emph{Findings of the Association for Computational Linguistics:
  ACL 2024}, pages 10471--10506, 2024.

\bibitem[Zheng et~al.(2025)Zheng, Fu, Hu, Cai, Ye, Lu, and
  Liu]{zheng2025deepresearcher}
Yuxiang Zheng, Dayuan Fu, Xiangkun Hu, Xiaojie Cai, Lyumanshan Ye, Pengrui Lu,
  and Pengfei Liu.
\newblock Deepresearcher: Scaling deep research via reinforcement learning in
  real-world environments, 2025.
\newblock URL \url{https://arxiv.org/abs/2504.03160}.

\end{thebibliography}

\clearpage
\appendix

 \section{Configuration Examples}
\label{app:configs}

This appendix collects the implementation details and configuration examples
referenced from \S\ref{sec:methodology}.  The details are separated from the
main text to keep the description of the progressive funnel compact.

\subsection{L1 State Windows and Short Circuits}

L1 distinguishes read-only, limited-write, system-interaction, and
high-danger tools; shell-like tools are additionally classified from command
content and their normalized effects.  D4 takes the maximum of two
session-local signals: (i) the number of recent high-risk events and (ii)
frequency anomalies.  Its high-risk horizon is configurable (300 seconds in
benchmark mode); events outside a finite horizon are evicted.  The default
frequency rules assign D4 \(=2\) for at least 10 calls to one tool within five
seconds, D4 \(\geq 1\) for at least 20 calls to one tool within 60 seconds, and
D4 \(\geq 1\) for at least 60 total calls within 60 seconds.  All horizons and
thresholds are operator-configurable.

Before composite scoring, a high-danger tool acting on a credential or system
target (SC-1), or a recognized high-danger command (SC-2), closes at
\textsc{Critical}; a pure read on a normal workspace target (SC-3) closes at
\textsc{Low} unless injection or later evidence invalidates that shortcut.
Effect-aware paths separately recognize a high-confidence equivalent of a
disabled capability (SC-4), an exact repeat of a previously denied effect
(SC-5), and a generated future-execution path associated with low-trust skill
evidence (SC-8).  The latter cases preserve or deepen review according to the
operating profile.  An unresolved script or wrapper is routed to semantic
review rather than being hard-blocked merely because a parser lacks evidence.

\begin{figure}[htbp]
\begin{maccode}[Resolved L1 High-Preset Parameters]{yaml}
{
  "preset": "high",
  "weights": {
    "max_d123": 0.40,
    "d4": 0.25,
    "d5": 0.15,
    "d6_multiplier": 0.70
  },
  "thresholds": {
    "medium": 0.50,
    "high": 1.20,
    "critical": 1.80
  }
}
\end{maccode}
\caption{Resolved L1 score parameters for the high-security benchmark preset.}
\label{fig:l1_policy}
\end{figure}

\subsection{L2 Rule Extension Interface}

\texttt{CS\_ATTACK\_PATTERNS\_PATH} selects an operator-provided YAML library;
without it, L2 loads the bundled 25-pattern library covering ASI01 goal
hijacking, ASI02 tool misuse and data exfiltration, ASI03 privilege abuse,
ASI04 supply-chain attacks, and ASI05 arbitrary code execution.  A pattern can
specify trigger logic, weighted regular expressions, false-positive filters,
risk escalation, and reference metadata, as illustrated in
Figure~\ref{fig:yaml_pattern}.

L2 separates rule production from enforcement through this versioned artifact
boundary.  The matcher parses and safe-compiles an operator-supplied pattern
snapshot and supports snapshot reload.  This decoupling also provides a clean
integration point for an external evolution engine: candidate rules can be
generated and validated outside the decision path, while only
operator-approved YAML snapshots are published to L2.  Such an extension
changes how rules are produced without transferring enforcement authority
away from the gateway.

\begin{figure}[htbp]
\begin{maccode}[YAML Attack Pattern]{yaml}
patterns:
  - id: "ASI02-001"
    category: "tool_misuse"
    risk_level: "critical"
    triggers:
      tool_names: ["bash", "shell", "exec"]
    detection:
      regex_patterns:
        - pattern: "curl.*--upload-file"
          weight: 8
    false_positive_filters:
      - type: "whitelist_path"
        paths: ["*/test_*", "*/ci/*"]
\end{maccode}
\caption{An excerpt of a customizable YAML attack pattern used by the L2 \texttt{RuleBasedAnalyzer}.}
\label{fig:yaml_pattern}
\end{figure}
\FloatBarrier

\subsection{L3 Skills and Tool Permissions}

L3 loads bundled YAML review manifests and can add operator-defined manifests
from \path{AHP_SKILLS_DIR}.  A manifest declares deterministic triggers
over risk hints, canonical tool names, and payload substrings; a reviewer
prompt and evaluation criteria; evidence, rubric, and tag metadata; a
per-skill \texttt{allowed\_tools} list; and an optional
\texttt{max\_tool\_calls}.  The registry ranks matching specialized skills and
falls back to \texttt{general-review} when none matches.

The tool boundary remains gateway-owned.  Every manifest allowlist is
validated against a fixed global read-only set.  At runtime, a requested tool
must appear in both the global set and the selected skill's allowlist, and the
effective call budget is capped by both analyzer and manifest budgets.
Consequently, a skill can specialize evidence collection and review criteria
but cannot add execution, mutation, installation, or network authority.

\begin{figure}[htbp]
\begin{maccode}[Illustrative L3 Review Skill]{yaml}
name: shell-risk-review
schema_version: clawsentry.l3_skill.v1
description: Review risky shell effects
enabled: true
priority: 10
triggers:
  risk_hints: [shell_execution]
  tool_names: [bash, shell]
  payload_patterns: ["curl | sh"]
required_evidence: [current_event, trigger]
allowed_tools: [read_trajectory, read_session_risk]
max_tool_calls: 4
severity_rubric:
  high: [destructive system effect]
output_tags: [shell_risk]
system_prompt: |
  Review bounded read-only evidence.
  Return JSON with risk_level, findings, confidence.
evaluation_criteria:
  - name: dangerous_effect
    severity: high
    description: Destructive or privilege-changing action
\end{maccode}
\caption{Schematic operator-defined L3 review skill aligned with the runtime
manifest schema. A skill specializes triggering, evidence inspection, and
review criteria while remaining within the gateway's global read-only tool
boundary.}
\label{fig:l3_skill}
\end{figure}
\FloatBarrier

\section{SkillsSafety Supporting Results}
\label{app:skillsafety}

SkillsSafety groups its 155 cases into six domains: RD1 contextual trust and
prompt manipulation, RD2 agent permission and authorization, RD3 execution,
runtime, and protocol, RD4 data boundaries and output exfiltration, RD5
memory, recovery, and audit, and RD6 knowledge-model, supply-chain, and
operational risks.

Table~\ref{tab:skillsafety_raw_rd} reproduces the per-domain raw baselines
published with the benchmark~\cite{jin2026skillsafetybench} for these same five
Work Agent configurations. We quote the published values rather than re-running
the undefended conditions, providing a fixed public reference point established
before the protected runs reported here.
Table~\ref{tab:skillsafety_counts}
then gives the auditable counts behind the protected operating points in
Table~\ref{tab:skillsafety_main}: AS denotes unsafe attack success and TS
successful completion of the benign user task, two outcomes a case realizes
independently.

\begin{table*}[t]
\centering
\begin{tabular}{@{}lcccccccc@{}}
\toprule
\textbf{Raw Work Agent} & \textbf{RD1} & \textbf{RD2} &
\textbf{RD3} & \textbf{RD4} & \textbf{RD5} & \textbf{RD6} &
\textbf{Overall ASR} & \textbf{TSR} \\
\midrule
Codex / GPT-5.4 & 64.0 & 52.0 & 30.8 & 51.9 & 46.2 & 23.1 & 44.5 & 38.7 \\
Codex / GPT-5.5 & 64.0 & 48.0 & 11.5 & 48.1 & 53.8 & 26.9 & 41.8 & 42.6 \\
Claude Code / GLM-5.1 & 36.0 & 44.0 & 19.2 & 44.4 & 38.5 & 19.2 & 33.5 & 37.4 \\
Claude Code / MiniMax-2.7 & 48.0 & 48.0 & 15.4 & 37.0 & 38.5 & 19.2 & 34.2 & 20.6 \\
Kimi CLI / Kimi-K2.5 & 76.0 & 72.0 & 26.9 & 55.6 & 50.0 & 19.2 & 49.7 & 31.0 \\
\bottomrule
\end{tabular}
\caption{Per-domain SkillsSafety raw baselines (\%), as published with the
benchmark~\cite{jin2026skillsafetybench}. RD columns and overall
are ASR (lower is better); TSR is benign task completion (higher is better).}
\label{tab:skillsafety_raw_rd}
\end{table*}

\begin{table*}[t]
\centering
\begin{tabular}{@{}lccccc@{}}
\toprule
\textbf{Protected Work Agent} & \textbf{N} & \textbf{AS} & \textbf{TS} &
\textbf{ASR}$\downarrow$ & \textbf{TSR}$\uparrow$ \\
\midrule
Codex / GPT-5.4 & 152 & 19 & 39 & 12.50 & 25.66 \\
Codex / GPT-5.5 & 153 & 23 & 46 & 15.03 & 30.07 \\
Claude Code / GLM-5.1 & 121 & 15 & 28 & 12.40 & 23.14 \\
Claude Code / MiniMax-2.7 & 154 & 14 & 24 & \textbf{9.09} & 15.58 \\
Kimi CLI / Kimi-K2.5 & 147 & 15 & 24 & 10.20 & 16.33 \\
\bottomrule
\end{tabular}
\caption{Counts behind the five protected SkillsSafety aggregate results
(\%). N is the evaluable-case denominator after applying the common exclusion
rule in \S\ref{sec:exp:setup}. All protected rows use
\texttt{gemini-3.5-flash} for the independent Gateway and FSPR review paths.}
\label{tab:skillsafety_counts}
\end{table*}

\section{SkillsSafety Fixed-Denominator TSR Comparison}
\label{app:tsr_analysis}

This appendix supports the analysis of \S\ref{sec:exp:tsr_analysis}, which
accounts for the protected SkillsSafety task-success margin by separating the
share attributable to the poisoning from the share attributable to enforcement.
Table~\ref{tab:tsr_rd} gives the six-domain counts behind
Figure~\ref{fig:tsr_poisoning}.

\paragraph{Pairing rule.}
The four conditions of \S\ref{sec:exp:tsr_analysis} hold the Codex/GPT-5.4 Work
Agent, the 60 sampled cases, and the scoring criterion fixed. Condition B is
constructed per case: each poisoned package is matched, through its original
task, to a clean SkillsBench~\cite{li2026skillsbench} package of the same class,
so the clean reference carries comparable task value rather than an arbitrary
substitute skill.

\paragraph{Denominator convention.}
Only the denominator differs from the other TSR experiments
(\S\ref{sec:exp:setup}). Partial, missing, and exceptional outcomes remain as
non-successes rather than being excluded, fixing every domain denominator at
ten. This keeps all four columns on a single comparison base, so a difference
between columns cannot arise from differing technical coverage.

\paragraph{Per-domain reading.}
The aggregate pattern is not driven by a single domain. Clean skills raise
full-task passes over the no-skill baseline in RD2, RD5, and RD6, and poisoning
lowers passes relative to the clean skill in all six domains. RD1 is the
sharpest case: 8/10 with a clean skill against 3/10 once the same class of
skill is poisoned, with ClawSentry disabled in both.
The enforcement margin from C to D is likewise concentrated rather than uniform:
D matches C in RD2 and RD4 and exceeds it in RD1, 5/10 against 3/10, where
refusing the package also prevented the derailment the package itself caused,
while RD3, RD5, and RD6 account for six differing cases between them, leaving a
net difference of four.

\paragraph{Why A is not the reachable reference for D.}
Reading D against condition A---the agent working alone at 43.3\%---presumes
that refusing a poisoned package restores the pre-session alternative. Two
properties of the setting, both visible in the per-domain counts, place D on a
different base. First, the decision is per-action inside a live session rather
than a pre-session choice of skill: by the time an action is denied the package
has been read and has shaped the plan, so refusal truncates the trajectory in
progress instead of substituting condition A's trajectory for it. RD1 shows the
mechanism operating in the favorable direction---5/10 for D against 3/10 for
C---because there the derailment the package caused is precisely what the block
prevented. Second, part of the C-to-D difference is the overblocking that
\S\ref{sec:exp:tsr_analysis} reports directly: the policy behind FSPR's
5.7--7.5\% clean false-block rate, measured there across two independent
review backends (\S\ref{sec:exp:fspr}), also declines some
benign actions here, and the unattended convention of resolving every
\textsc{Defer} as \textsc{Block} (Appendix~\ref{app:defer}) places D at the most
conservative point of that curve rather than at its attended one. RD3 and RD5,
accounting for three and two of those cases, are where this concentrates.

\begin{table*}[t]
\centering
\small
\setlength{\tabcolsep}{4pt}
\begin{tabularx}{\textwidth}{@{}>{\raggedright\arraybackslash}Xccccc@{}}
\toprule
\textbf{Risk domain} & \textbf{Cases} & \textbf{A: no skill} &
\textbf{B: clean} & \textbf{C: poisoned} & \textbf{D: +CS} \\
\midrule
RD1: contextual trust / prompt manipulation & 10 & 9/10 & 8/10 & 3/10 & 5/10 \\
RD2: agent permission / authorization & 10 & 2/10 & 4/10 & 2/10 & 2/10 \\
RD3: execution / runtime / protocol & 10 & 4/10 & 4/10 & 3/10 & 0/10 \\
RD4: data boundary / output exfiltration & 10 & 5/10 & 5/10 & 4/10 & 4/10 \\
RD5: memory / recovery / audit & 10 & 4/10 & 5/10 & 4/10 & 2/10 \\
RD6: knowledge model / supply chain / operations & 10 & 2/10 & 4/10 & 3/10 & 2/10 \\
\midrule
\textbf{Total} & \textbf{60} & \textbf{26/60 (43.3\%)} &
\textbf{30/60 (50.0\%)} & \textbf{19/60 (31.7\%)} &
\textbf{15/60 (25.0\%)} \\
\bottomrule
\end{tabularx}
\caption{Per-domain full-task-success counts for the fixed-denominator
four-condition comparison. A--C carry no ClawSentry component; D applies the
hard-block policy. The security benefit of D is evaluated by ASR elsewhere;
TSR here characterizes the poisoned-package utility operating point.}
\label{tab:tsr_rd}
\end{table*}

\section{Controlled Security Comparisons}
\label{app:fspr_modes}

\subsection{FSPR Response Modes}

Once the package scanner distrusts a skill, the response is itself a design
decision. This study compares three responses on a 60-case SkillsSafety subset
sampled at random within risk domains (ten per domain, fixed seed), with
Claude Code (MiniMax-M3) as Work Agent and \texttt{gemini-3.5-flash}
as ClawSentry reviewer. Only the response to the FSPR finding changes.
\emph{Hard block} refuses execution of the package outright. \emph{Warning}
admits the package but injects a caution into its documentation naming the
suspected risk area. \emph{Feedback} admits the package and delivers a runtime
safety note through a host hook, as if raised by the user.

Beyond the hard-block versus warning contrast reported in
\S\ref{sec:exp:ablation}, this fixture puts hard block against the strongest
soft response available: relative to runtime feedback, hard block is 2.06 points
lower in ASR and 1.10 points higher in TSR
(Figure~\ref{fig:controlled_comparisons}(a)). Hard block therefore leads whether
it is compared with the weakest soft response or the strongest.

\paragraph{Why soft responses are structurally weaker.}
\S\ref{sec:exp:ablation} locates the gap in where a soft response sits relative
to a context the attacker has already shaped. Two further effects compound that
ordering problem. \emph{Precision:} a caution able to name the exact defect
would already require the deep analysis the soft mode was meant to avoid, so it
stays coarse and the model can reasonably judge the package benign.
\emph{Alignment dependence:} once a late notice competes with package contents
the model has already read, the outcome is settled by the Work Agent's own
safety alignment, so a soft mode's strength varies with the model while
admission refusal is a property of the policy.

The shape of the design space governs how the comparison should be read. The
three responses differ in how tightly they couple to the decision
itself---warning acts at the document level, feedback on a runtime channel
adjacent to the decision, and hard block at the decision boundary---and ASR
orders with that coupling. Feedback additionally depends on the
host surface: several agent surfaces are audit-only and deliver no visible
runtime response, so a visible-feedback arm is well defined only where the
surface supports it, and a default policy resting on it would not carry
across integrations. A fourth and heavier option---sanitizing the package into a
clean version so that its benign functionality survives---is deliberately placed
outside the runtime path: it is costly and can introduce new faults, so we treat
it as the lifecycle direction identified in \S\ref{sec:exp:tsr_analysis} rather
than as a runtime response mode.

\subsection{Matched Defense Baseline}

We compare ClawSentry with the open-source \texttt{claude-guardrails}
PreToolUse hooks~\cite{dwarvesf2026guardrails} and a raw configuration. All
three use Claude Code with MiniMax-M3 on the same fixed 60-case SkillsSafety
subset; only the defense changes. The baseline is a set of local rule hooks
with no model-backed review, evaluated at the commit pinned in our artifact.

ClawSentry's 8.62\% ASR is 48.52 percentage points below
\texttt{claude-guardrails} and 49.71 points below raw
(Figure~\ref{fig:controlled_comparisons}(b)). The additional reading this
fixture supplies is the size of the baseline's own movement: the rule hooks shift
ASR by 1.19 points from the undefended configuration, which places the
limitation in coverage rather than in strictness---an action semantically
equivalent to one the pattern set does block, expressed in a form the shipped
patterns do not enumerate, reaches execution unchanged. TSR is reported on a
separate utility axis, and the three arms admit attack traffic that differs by up
to 49.71 points, so the task-success columns describe three different operating
points rather than a single exchange rate between the two axes.

\subsection{Component Contributions}

This experiment fixes Codex/GPT-5.4, the \texttt{gemini-3.5-flash} reviewer, a
fixed 60-case SkillsSafety subset, and all other policy settings. It disables
FSPR and anti-bypass individually.

Figure~\ref{fig:controlled_comparisons}(c) places the two removals reported in
\S\ref{sec:exp:ablation} beside their TSR operating points. The accompanying TSR
movement follows the account in \S\ref{sec:exp:tsr_analysis}: relaxing either
boundary readmits residual utility from packages that remain poisoned, which is
precisely the exchange the full configuration declines.

\section{Reviewer Model Replaceability}
\label{app:reviewer}

We fix the Codex/GPT-5.4 Work Agent, a 60-case SkillsSafety sample drawn at
random under a fixed seed, ClawSentry policy, and code snapshot, then replace
only the safety reviewer
model. Each configuration supplies both Gateway L2/L3 runtime review and FSPR
first-use review, but the two paths retain separate contracts and counting
units. The experiment therefore measures a system-level reviewer substitution,
not an isolated L3 or FSPR model ablation. Its purpose is not to rank reviewer
models but to test whether the safety boundary is a property of the harness
rather than of one model family.

\begin{table*}[t]
\centering
\begin{tabular}{@{}lccccc@{}}
\toprule
\textbf{Reviewer model} & \textbf{N} & \textbf{AS} & \textbf{TS} &
\textbf{ASR}$\downarrow$ & \textbf{TSR}$\uparrow$ \\
\midrule
\texttt{gemini-3.5-flash} & 55 & 7 & 14 & 12.73 & 25.45 \\
\texttt{minimax-m2.7} & 54 & 6 & 15 & \textbf{11.11} & \textbf{27.78} \\
\texttt{deepseek-v4-pro} & 58 & 10 & 13 & 17.24 & 22.41 \\
\texttt{minimax-m3} & 57 & 7 & 14 & 12.28 & 24.56 \\
\texttt{glm-5.1} & 58 & 10 & 10 & 17.24 & 17.24 \\
\bottomrule
\end{tabular}
\caption{System-level reviewer substitution on 60 attempted SkillsSafety cases
(\%). N is the evaluable-case denominator. The reviewer model supplies safety
analysis only; the Work Agent is Codex/GPT-5.4 in every row.}
\label{tab:reviewer}
\end{table*}

Table~\ref{tab:reviewer} gives the per-reviewer counts behind the ASR--TSR band
of \S\ref{sec:exp:ablation}. Three of the five reviewers fall within 1.7 points
of one another---\texttt{minimax-m2.7} at 11.11\%, \texttt{minimax-m3} at
12.28\%, and the main-experiment default \texttt{gemini-3.5-flash} at
12.73\%---and the two highest values coincide at 17.24\% across different
provider families, so the residual spread tracks neither a single vendor nor the
reviewer chosen for the main experiments. This is the decoupling the
architecture intends: safety capability can be supplied to a Work Agent by an
independent reviewer, and substituting that reviewer moves the operating point
within an enforcement chain the harness continues to maintain.

\section{Gateway Funnel and First-Use Review Volume}
\label{app:cost}

Table~\ref{tab:cost} uses the same five reviewer substitutions as
Appendix~\ref{app:reviewer}; main-text Figure~\ref{fig:portability_funnel}(b)
summarizes the resulting aggregate funnel and
Figure~\ref{fig:portability_funnel}(c) the
per-reviewer total Gateway share.
Counts are module events rather than unique
cases: one case may produce multiple tool actions, reviews, or follow-up
guards. Gateway review is the sum of L2 semantic-review calls and L3
deep-review requests. FSPR is a separate package-level first-use path, and
anti-bypass is session-level follow-up state; neither is folded into the
Gateway ratio.

\begin{table*}[t]
\centering
\begin{tabular}{@{}lrrrrrrr@{}}
\toprule
\textbf{Reviewer model} & \textbf{L1} & \textbf{L2} & \textbf{L3} &
\textbf{L2+L3} & \textbf{Gateway/L1} & \textbf{FSPR} &
\textbf{Anti-bypass} \\
\midrule
\texttt{gemini-3.5-flash} & 1{,}646 & 233 & 22 & 255 & 15.5\% & 127 & 112 \\
\texttt{minimax-m2.7} & 1{,}591 & 210 & 16 & 226 & 14.2\% & 134 & 77 \\
\texttt{deepseek-v4-pro} & 1{,}661 & 182 & 28 & 210 & 12.6\% & 165 & 86 \\
\texttt{minimax-m3} & 1{,}669 & 220 & 26 & 246 & 14.7\% & 124 & 105 \\
\texttt{glm-5.1} & 1{,}743 & 191 & 22 & 213 & 12.2\% & 139 & 103 \\
\midrule
\textbf{Total} & \textbf{8{,}310} & \textbf{1{,}036} &
\textbf{114} & \textbf{1{,}150} & \textbf{13.8\%} &
\textbf{689} & \textbf{483} \\
\bottomrule
\end{tabular}
\caption{Observed request and trigger volume across five reviewer
substitutions and 300 provider--case executions. L3 denotes Gateway L3 only;
FSPR remains a separate first-use package review.}
\label{tab:cost}
\end{table*}

Decomposing the aggregate ratio of \S\ref{sec:exp:cost}: L2/L1 is 12.5\% and
L3/L2 is 11.0\%, and per reviewer L2/L1 stays within 11.0--14.2\% and L3/L1
within 1.0--1.7\%. FSPR runs 2.30 first-use package reviews per
provider--case, counted outside the Gateway ratio.

The stability of these ratios across reviewers matters for the cost argument of
\S\ref{sec:exp:cost}: the funnel shape is set by the policy and the traffic,
not by which model answers the review, so an operator can substitute reviewers
without re-tuning the escalation budget. The 483 anti-bypass triggers are
counted separately for the same reason---they are session state consulted at the
end of the decision chain, not additional model-backed requests. What these
counts establish is the quantity the architecture itself controls, namely
review-request volume; per-request latency, token consumption, and pricing
remain properties of the reviewer backend an operator chooses.

\section{Unattended Defer Convention}
\label{app:defer}

\textsc{Defer} represents a boundary or uncertain request that should receive
assisted review: in an interactive deployment, a human inspects the evidence
and approves the action when appropriate. The benchmark runs without an
operator and must terminate deterministically, so it enforces every pre-action
\textsc{Defer} as \textsc{Block} (\S\ref{sec:exp:setup}). This mapping
can conservatively refuse legitimate work that an interactive reviewer would
approve, so the reported protected TSR is a lower bound on the attended
operating point. The overblocking measured under benchmark conditions therefore
includes the cost of this convention and characterizes the unattended operating
point rather than the false-block behavior of an attended deployment. Every
reported ASR and TSR uses the unattended mapping throughout; no unobserved human
decision is imputed.

\section{Task-Artifact Trust in the SkillsSafety Evaluation}
\label{app:scope_trust}

SkillsSafety cases run as root inside containers, so their verifier-mandated
data and output locations are absolute paths under roots such as \texttt{/app}
and \texttt{/root}---a surface that target-sensitivity rules rightly treat as
sensitive.  The benchmark harness therefore acts as a task-runner grantor of the
interface in \S\ref{sec:method:scope}, declaring each case's mandated
artifact paths exactly as a user or job scheduler assigns a job its
input and output directories; the gateway applies the same deterministic
profile evaluation as for any other grantor, with no benchmark-specific code
path.  Two limits keep the declaration from widening the attack surface.
Declarations are accepted only from benchmark-control channels---runner and
verifier metadata, cross-checked between the task instruction and the verifier
tests---and they cover only task-data reads and task-output writes.  Verifier and
oracle paths are never declared, and every other effect flows through the
unchanged pipeline, so no attack-relevant behavior receives an allowance.

\end{document}